\documentclass[12pt,preprint]{aastex}

\shorttitle{NLTE 1.5D Red Giants}
\shortauthors{Young \& Short}

\begin{document}


\title{NLTE 1.5D Modeling of Red Giant Stars}


\author{Mitchell. E. Young}
\affil{Department of Astronomy \& Physics and Institute for Computational Astrophysics, Saint Mary's University,
    Halifax, NS, Canada, B3H 3C3}
\email{myoung@ap.smu.ca}

\author{C. Ian Short}
\affil{Department of Astronomy \& Physics and Institute for Computational Astrophysics, Saint Mary's University,
    Halifax, NS, Canada, B3H 3C3}
\email{}


\begin{abstract}

Spectra for 2D stars in the 1.5D approximation are created from synthetic spectra of 1D non-local thermodynamic equilibrium (NLTE) spherical model atmospheres produced by the PHOENIX code. The 1.5D stars have the spatially averaged Rayleigh-Jeans flux of a K3-4 III star, while varying the temperature difference between the two 1D component models ($\Delta T_{\mathrm{1.5D}}$), and the relative surface area covered. Synthetic observable quantities from the 1.5D stars are fitted with quantities from NLTE and local thermodynamic equilibrium (LTE) 1D models to assess the errors in inferred $T_{\mathrm{eff}}$ values from assuming horizontal homogeneity and LTE. Five different quantities are fit to determine the $T_{\mathrm{eff}}$ of the 1.5D stars: UBVRI photometric colors, absolute surface flux SEDs, relative SEDs, continuum normalized spectra, and TiO band profiles. In all cases except the TiO band profiles, the inferred $T_{\mathrm{eff}}$ value increases with increasing $\Delta T_{\mathrm{1.5D}}$. In all cases, the inferred $T_{\mathrm{eff}}$ value from fitting 1D LTE quantities is higher than from fitting 1D NLTE quantities and is approximately constant as a function of $\Delta T_{\mathrm{1.5D}}$ within each case. The difference between LTE and NLTE for the TiO bands is caused indirectly by the NLTE temperature structure of the upper atmosphere, as the bands are computed in LTE. We conclude that the difference between $T_{\mathrm{eff}}$ values derived from NLTE and LTE modelling is relatively insensitive to the degree of the horizontal inhomogeneity of the star being modeled, and largely depends on the observable quantity being fit.

\end{abstract}

\keywords{stars: atmospheres, fundamental parameters, late-type --- methods: data
analysis --- techniques: photometric, spectroscopic}

\section{INTRODUCTION}

\subsection{Red Giant Stars}

Red giant stars rank among the brightest stars in the Galaxy, being generally 
much brighter in the visible band than main sequence stars of the same spectral 
type. But this only partially accounts for their brightness as large portion of 
their flux is emitted in the near-IR. This, combined with their enormous surface 
areas, gives them such large luminosities that they are easily observable even in 
very remote stellar populations. Red giants grant us a tool for probing nearly 
all regions of the Galaxy using a common indicator, a feat unparalleled by most 
other types of stars. 

\paragraph{}

Because many red giants are low to intermediate mass stars that have evolved
beyond the main sequence, generally found in older stellar populations,
their abundances can be indicators of early Galactic chemical evolution.
For example, by comparing observations of Galactic bulge giants with
those of giants located in the thin and thick disks and halo, it has
been shown that the bulge likely experienced similar formation timescales,
chemical evolution histories, star formation rates and initial mass
functions as the local thick disk population \citep{melendez08,alves-brito10}.
Bulge and disk giants show some differences in their chemical abundances,
with the bulge giants showing a higher relative abundance of select
elements than the disk giants, suggesting more rapid chemical enrichment,
possibly by ejecta from supernovae of Types Ia and II \citep{cunha06}.
Observations of red clump giants in the bulge have also produced additional
evidence of a central bar \citep{stanek97}, with their
apparent visual magnitudes being brighter at some Galactic latitudes
than others.

\paragraph{}

The tip of the red giant branch (TRGB) can also be used as a standard candle to 
determine distances to nearby galaxies. The distance moduli obtained from the
$I$ pass-band magnitude of the TRGB are comparable with those from
primary distance indicators like Cepheids and RR Lyraes \citep{makarov06},
and in some cases even suggest a reevaluation of the metallicity dependence
and zero point calibration of the Cepheid distance scale \citep{salaris98,rizzi07}.
The TRGB method even has advantages over other distance determinations
like those of Cepheids and RR Lyraes: $\left(1\right)$ the TRGB method
requires much less telescope time than variable stars; $\left(2\right)$
the $I$ magnitude of the TRGB is insensitive to the variation of
metallicity for $\mathrm{[Fe/H]}<-$ 0.7; and $\left(3\right)$ the
TRGB suffers less from extinction problems than Cepheids, which are
in general located in star-forming regions \citep{lee93}.

\subsection{Modeling Stellar Atmospheres\label{intromodel}}

Much of what is now known about all types of stars comes from fitting
the predicted quantities from atmospheric models to observations.
Estimates of the solar chemical abundances come from fitting synthetic
spectral line profiles and equivalent widths to those observed in
the Sun \citep{ross76,asplund09,caffau11}. Calibrations of
stellar parameters, such as $g$, $M_V$, $T_{\mathrm{eff}}$,
$L$, and $R_{0}$, for different spectral types are found from
fitting models \citep{martins05}. Beyond studying single stars,
model atmospheres can be used to determine qualities of larger structures
as well. The age-metallicity and color-metallicity relations of globular
clusters can be determined from the abundances of individual red giants
within the clusters \citep{pilachowski83,carretta97,carretta10}. 

\paragraph{}

In this work, we will explore the limitations of two of the simplifying 
assumptions of atmospheric modeling: horizontal homogeneity and local 
thermodynamic equilibrium (LTE). Both of these assumptions have been generally 
adopted because they are more computationally practical than the alternatives, 
requiring less time and fewer resources to arrive at a result. By comparison, 
horizontal inhomogeneity requires model atmospheres to be calculated in two or 
three geometric dimensions (2D or 3D models) instead of just one, and non-LTE 
(NLTE) requires that each level population be computed in statistical equilibrium 
(SE) using iterative processes.

\paragraph{}

However, horizontal homogeneity and LTE both limit how realistic a
model can be. Simulations of red giant atmospheres performed in 3D
have confirmed that turbulent surface convection causes horizontal
inhomogeneities to form \citep{collet07,kucinskas13b}, such
as visually observable surface features like solar granulation \citep{mathur11,tremblay13}.
These features are known to lead to detectable effects, such as altering predicted
line strengths and shapes and, thus, inferred elemental abundances
\citep{collet08,collet09,dobrov13,hayek11,kucinskas13a,mashonkina13}.
For 3D models of red giant atmospheres, whose modeling parameters
span the ranges of 3600 K $\leq T_{\mathrm{eff}}\leq$ 5200 K, 1.0
$\leq\mathrm{log\,}g\leq$ 3.0, and -3.0 $\leq\mathrm{[Fe/H]}\leq$
0.0, granules have been shown to span a range of sizes from as small
as on the order of 10$^{8}$ cm to as large as 2 $\times$ 10$^{12}$
cm, with the majority on the order of 10$^{11}$ cm \citep{collet07,chiavassa10,hayek11,ludwig12,magic13b,tremblay13}.
The cooler stars and stars with lower values of $\mathrm{log}\, g$
generally display larger features. For the same set of 3D models,
the root mean square (RMS) temperature variation among these features at optical depth unity is 
usually in the range of $\sim$ 2 to 5 $\%$, or $\sim$ 200 to 300 K for the parameters 
listed above, although variations can reach 2000 K between the hottest and 
coolest areas \citep{collet08,collet09,kucinskas13a,kucinskas13b,ludwig12,magic13a,magic13b,samadi13,tremblay13}.
Most of the 3D models reported in the literature have $T_{\mathrm{eff}}$ $\approx$
4500 K, with $\Delta T$ varying by $\sim$ 200 K.

\paragraph{}

For stars exhibiting horizontal inhomogeneities, where the temperature varies across 
the features, $T_{\mathrm{eff}}$ is no longer a well defined quantity. By definition,
$T_{\mathrm{eff}}$ is derived from the Stefan-Boltzmann law, where the bolometric 
luminosity ($L_{\mathrm{bol}}$) of a star is proportional to the fourth power 
of its $T_{\mathrm{eff}}$ ($T_{\mathrm{eff,S-B}}$). The $T_{\mathrm{eff,S-B}}$ of a horizontally inhomogeneous 
star may be similarly found by summing the $L_{\mathrm{bol}}$ of each of the 
inhomogeneous components, weighted by their relative surface coverage, and taking 
the fourth root of the result. This quantity is defined 
empirically by intrinsic properties of real or model stars and is independent of 
fitting models to observable quantities. Alternatively, another model fitting 
independent $T_{\mathrm{eff}}$ can be defined from the long wavelength tails of 
stellar spectra. Because flux has a linear dependence on $T_{\mathrm{eff}}$ in 
the Rayleigh-Jeans limit, an estimate of the $T_{\mathrm{eff}}$ can be made from 
measuring the absolute surface flux of the R-J tail. This dependence is used by 
the Infrared Flux Method \citep{ramirez05} to determine $T_{\mathrm{eff}}$ from 
the R-J tails of spectra ($T_{\mathrm{eff,R-J}}$).

\paragraph{}

The key idea for a horizontally inhomogeneous star having the same 
$T_{\mathrm{eff,S-B}}$ as a horizontally homogeneous star, is that the SEDs will 
differ based on the variation of the modeling parameters, such as $T_{\mathrm{eff}}$, 
across the inhomogeneous surface. For example, using the Planck function,
\begin{equation}
B_{\lambda}(T_{\mathrm{eff}})=\frac{2hc^{2}}{\lambda^{5}}\left(e^{hc/\lambda k_{B}T_{\mathrm{eff}}}-1\right)^{-1},
\end{equation}
to describe the shape of a stellar continuum illustrates the issue directly \citep{uitenbroek11}.
A star with a range of differing temperatures across the surface does not have a 
directly obvious value of $T_{\mathrm{eff}}$ that should be used, and 
because of the equation's non-linear dependence on $T_{\mathrm{eff}}$, this is an 
important question. The higher temperature material will contribute 
disproportionately more flux than the lower temperature material, with a 
dependence on wavelength, and will alter the spectrum from that of a 
horizontally homogeneous star of the same $T_{\mathrm{eff,S-B}}$ 
accordingly. 

\paragraph{}

The most noticeable departure of LTE models from observed stars comes
from comparing computed and observed spectral features and SEDs. For
1D models of red giants, NLTE models have been shown to be
more accurate than LTE models in predicting the overall monochromatic
flux ($F_{\lambda}$) levels of SEDs and strengths of individual
spectral lines, with the notable exceptions of molecular absorption
bands and the near-UV band flux \citep{short03,short06,short09,bergemann13}.
Calculating the molecular level populations in NLTE is computationally
demanding, and is not handled in many atmospheric modeling codes.
Both LTE and NLTE models over-predict the near-UV $F_{\lambda}$ levels
of cool red giants. In the case of the near-UV $F_{\lambda}$ levels,
NLTE models are worse in the over-prediction than their LTE counterparts.
The NLTE effects of Fe group elements on the model structure and $F_{\lambda}$
distribution have been shown to be much more important for predicting
a SED than the NLTE effects of all the light metals combined, and
serve to substantially increase the near-UV $F_{\lambda}$ levels
as a result of NLTE Fe I overionization \citep{short09}. The magnitude of this 
effect has been shown to be inversely proportional to the completeness of the 
Fe I atomic model used in the atmospheric modeling \citep{mashonkina11}, discussed in 
detail in Section \ref{PHOENIX}. These failures of 1D NLTE models to predict 
observable quantities may be, in part, related to the exclusion of horizontal
inhomogeneities in the models.

\subsection{Present Work \label{sec:Present-Work}}

Our primary goal is investigating the relation between errors introduced
in determining a star's $T_{\mathrm{eff}}$ from assuming horizontal
homogeneity and those introduced by assuming LTE. We look to determine how 
distinguishable 3D hydro and NLTE effects are when looking at low resolution 
diagnostics (photometric colors and overall SEDs). The error inherent
in assuming LTE is expected to remain approximately constant for different
levels of horizontal inhomogeneity, as any changes in a LTE spectrum
caused by the inhomogeneities should also be represented to a similar
degree in the corresponding NLTE spectrum, as the relative difference
in the flux distribution between LTE and NLTE should remain roughly
constant within the $T_{\mathrm{eff}}$ range studied in this work. 

\paragraph{}

Section \ref{methods} outlines the parameters used for the 1D NLTE and LTE model 
grids, and the methods used in creating and processing the 1.5D NLTE spectra.  
Section \ref{Results} presents the results of fitting 1D SEDs and spectra to the 
1.5D SEDs and spectra. Section \ref{summary} gives a brief summary and discussion of the 
results.

\section{1D AND 1.5D ATMOSPHERIC MODELS \label{methods}}

In general, the calculation of a 2D model atmosphere requires
a very substantial increase in computational effort over that needed
for 1D models. However, there are two regimes of validity for 
which the situation simplifies; these are often referred to as 1.5D models.
The valid regimes are either 1) when the geometric scale of the inhomogeneities is so small
that they are very optically thin, and every emergent ray may be assumed 
to have fully sampled both warm and cool components of the atmosphere many times, 
or 2) that the inhomogeneities are large enough that they are optically thick to
their own radiation so that the flux is similar to that of two spatially unresolved stars.
Regardless, both regimes require that the inhomogeneities be small enough and randomly 
distributed so that there is no correlation between component $T_{\mathrm{eff}}$ values and 
limb darkening. By assuming the average size of surface features to be that of 3D hydro 
modeled red giant granules, we satisfy both conditions of the larger inhomogeneity
regime.

\paragraph{}

We construct our 1.5D SEDs by taking the linear average of two 1D NLTE SEDs. 
Two-dimensional information, such as the difference in temperature
among features or the relative portion of the stellar surface covered
by the different features, can be approximated by choosing the parameters
of the 1D components being averaged, and the averaging weights. These
1.5D NLTE SEDs and derived colors, approximating 2D effects, are treated as 
artificial ``observations'' of real stars that are fitted with a library of
NLTE and LTE 1D trial spectra for inferring the $T_{\mathrm{eff}}$ of the 1.5D
stars. A grid of 50 1D stellar model atmospheres and corresponding
synthetic spectra was produced for this purpose using the PHOENIX
code.

\subsection{PHOENIX \label{PHOENIX}}

\begin{figure}
\plotone{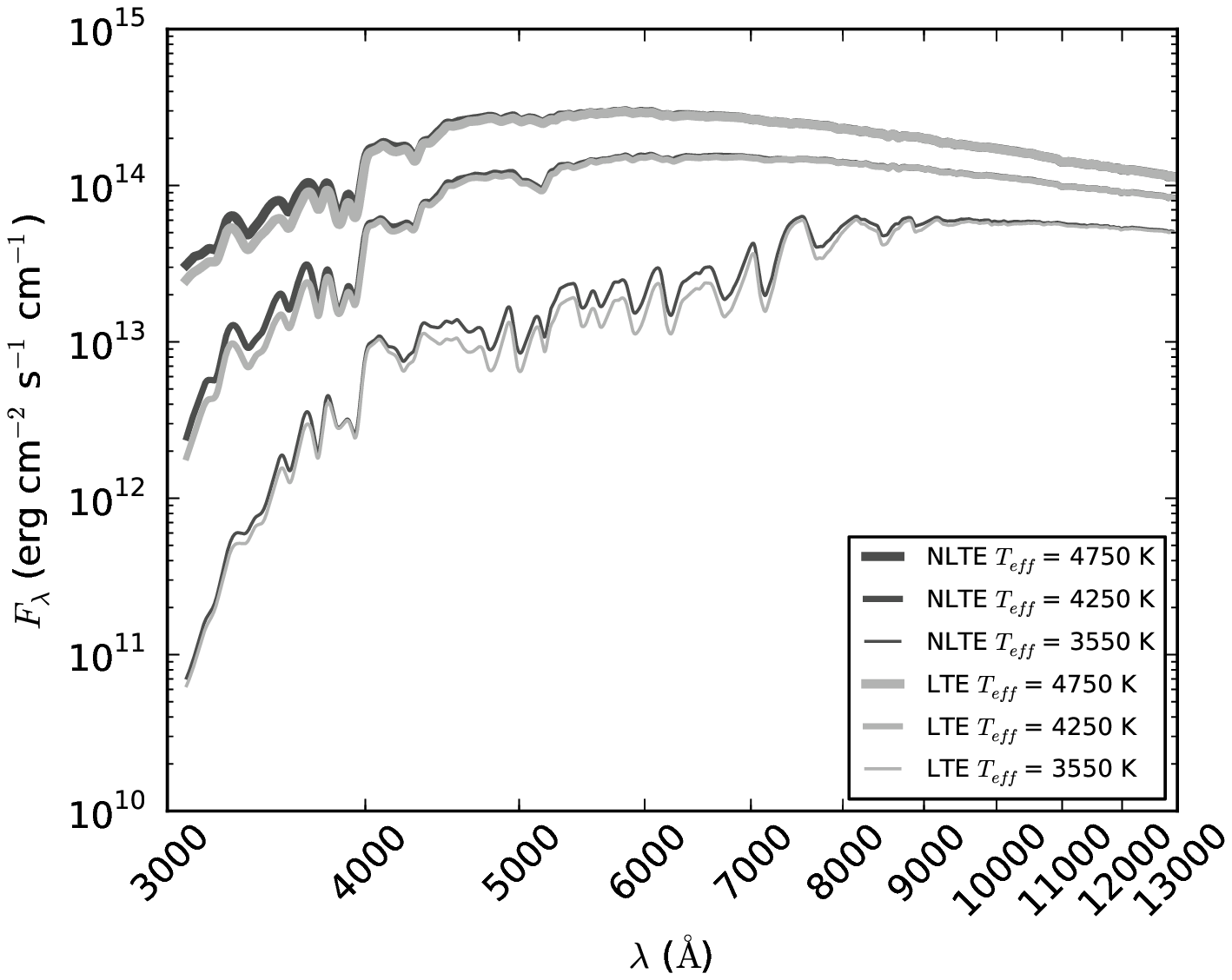}
\caption{PHOENIX LTE and NLTE synthetic spectra generated from the hottest and coolest grid models and the 1D model with the same $T_{\mathrm{eff,R-J}}$ (see Section \ref{intromodel} as the 1.5D spectra. The high resolution spectra generated by PHOENIX were convolved with a FWHM = 50 $\textrm{\AA}$ Gaussian kernel, and the resultant spectra are displayed. The LTE spectra with $T_{\mathrm{eff}}$ = 4250 K and 4750 K are mostly hidden behind their NLTE counterparts.\label{fig:exampleSEDs}}
\end{figure}

The PHOENIX code can be used to model NLTE atmospheres and spectra
of stellar objects throughout the H-R diagram \citep{baron98,hauschildt97,brott05}. 
PHOENIX version 15 is utilized for all modeling calculations in this work.
It is important to include NLTE effects in model atmospheres and synthetic 
spectra. Together, they can lead to a difference in the blue and near-UV bands 
flux of up to 50\% over corresponding PHOENIX LTE models of red giant stars as a result 
of NLTE Fe I overionization \citep{short03,short09}, displayed in Fig. \ref{fig:exampleSEDs}. 
It is also understood that NLTE radiative equilibrium in these stars is 
model-dependent and is known to depend on the completeness of the atomic model of 
Fe I and the completeness and accuracy of the collisional cross-sections for Fe I 
ionization. While the free electron collisional cross-sections for Fe I used by 
PHOENIX are robust, it does not take into account H collisions, and it has been 
cautioned that this may cause PHOENIX to overestimate the effects of Fe I overionization 
\citep{asplund05}. Additionally, it was shown by \citet{mashonkina11} that using 
a more complete Fe I atomic model than that used by PHOENIX will reduce overionization 
effects by providing more Fe I high energy excited states to facilitate 
recombinations from Fe II. They also found that the greater the number of energy 
levels within $\Delta E$ = k$T_{\mathrm{eff}}$ of the ground state ionization energy ($\chi_{\mathrm{ion}}$), 
the more accurate the NLTE ionization equilibrium solution. The number of energy 
levels in our model Fe I atom within $\Delta E$ = k$T_{\mathrm{eff}}$ of 
$\chi_{\mathrm{ion}}$ ranges from one to four as $T_{\mathrm{eff}}$ increases 
from 4250 ($T_{\mathrm{eff,R-J}}$ of our 1.5D spectra) to 4750 K (our hottest 1.5D 
component). This has been found to be too few to accurately compute the NLTE 
recombination rate, and we expect to over-estimate the NLTE Fe I overionization. 
However, we note that the 1.5D 'target' spectra and the 1D library of fitted NLTE 
spectra were computed with the \textit{same} Fe I atomic model. Therefore, we 
expect the difference between fitted 1D NLTE $T_{\mathrm{eff}}$ values and 
$T_{\mathrm{eff,S-B}}$ to be dominated by horizontal inhomogeneity effects. We 
also note that the magnitude of any difference in inferred $T_{\mathrm{eff}}$ 
value between fitting with 1D NLTE spectra and 1D LTE spectra will be over estimated, 
especially for any diagnostics involving wavebands bluer than the V filter, where the 
overionization effects are dominant. Therefore we concern ourselves primarily with 
how this difference depends on the degree of inhomogeneity in the 1.5D spectra being 
fit, and ignore the magnitude. 

\paragraph{}

The stellar parameters adopted for our $T_{\mathrm{eff,R-J}}$ equivalent 1D star are $T_{\mathrm{eff}}$
= 4250 K, log $g$ = 2.0, and {[}M/H{]} = -0.5 with the alpha elements,
from O to Ti, enhanced by {[}A/H{]} = +0.3 dex. The NLTE models were
constructed by treating the 20 atomic species listed in Table \ref{tab:nlte species list}
as NLTE species when calculating energy level populations, all of
them in the neutral state and most in the singly ionized state as
well \citep{short05}. PHOENIX does not calculate level populations for molecular
species in NLTE; molecules are treated in LTE for NLTE models and spectra.

\paragraph{}

\begin{figure}
\plotone{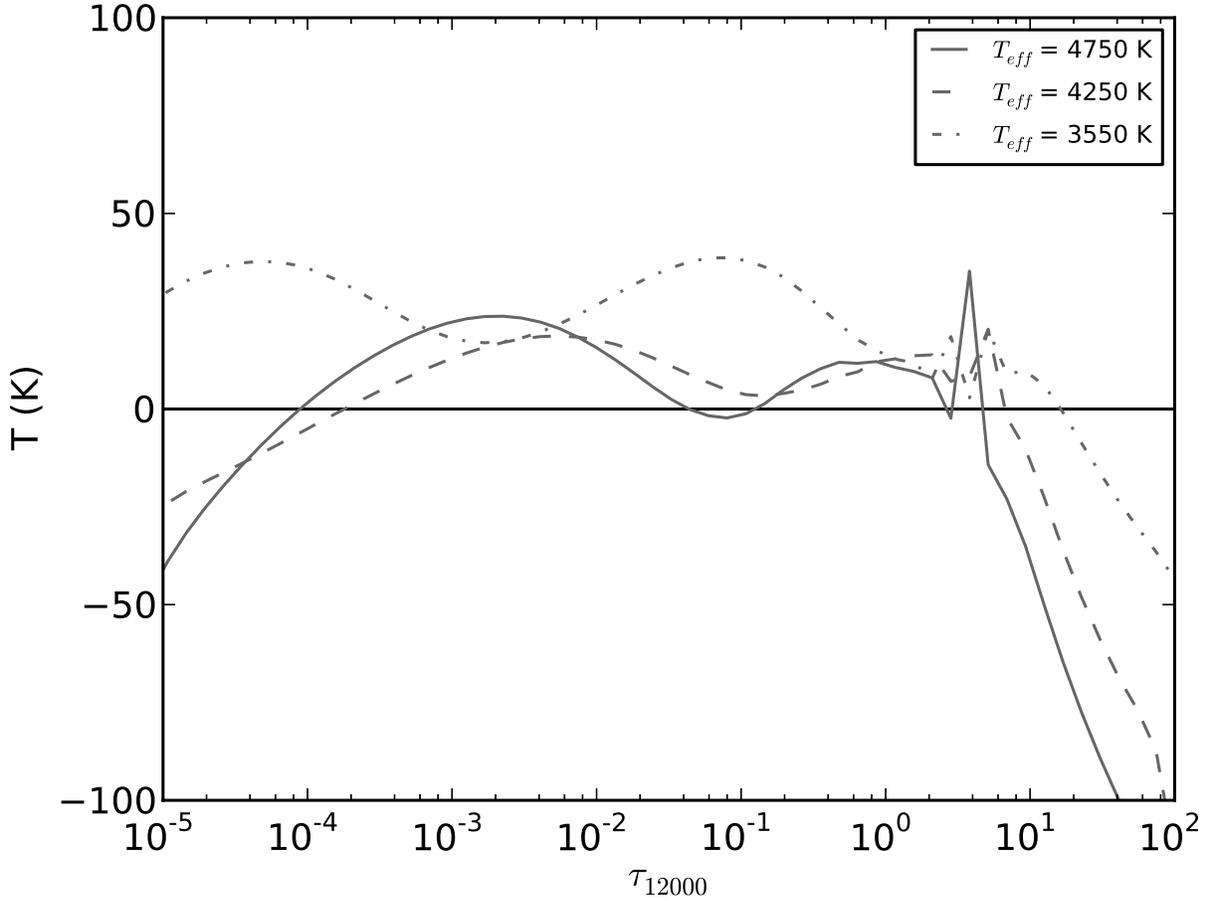}
\caption{$T_\mathrm{NLTE}$($\tau$) -- $T_\mathrm{LTE}$($\tau$) for PHOENIX atmospheric models as functions of the monochromatic continuum optical depth at 12000 $\textrm{\AA}$ representing the same three models as in Fig. \ref{fig:exampleSEDs}.\label{fig:exampleTtau}}
\end{figure}

NLTE radiative equilibrium is complex in that any given transition
may either heat or cool the atmosphere with respect to LTE, depending
on how rapidly $\tau_{\lambda}$ increases inward at the transition
wavelength, whether the transition is located in the Wien or Rayleigh-Jeans
regime, and whether the transition is a net heater or cooler in LTE
with respect to the gray atmosphere \citep{short12}. For our $T_{\mathrm{eff}}$
range , all of the NLTE models exhibit a surface cooling effect in
the outermost layers of the atmosphere. This is displayed in Fig.
\ref{fig:exampleTtau} as a lower temperature in the NLTE models than
in the LTE ones for a given $\tau_{12000}$. For the majority of the
upper atmosphere, between $\tau_{12000}\approx$ 10$^{-4}$ and 10$^{-1}$,
the opposite is seen, with the NLTE models having a higher temperature
than the LTE ones.

\paragraph{}

To prepare the grid of models, we varied the $T_{\mathrm{eff}}$ value 
from 3550 K to 4750 K, with $\Delta T_{\mathrm{eff}}$ = 50 K, and
produced both LTE and NLTE models. Upon convergence of a model structure, two 
different spectra were synthesized for each model, the fully line blanketed SED 
and the pure continuum SED for use in normalizing the flux in the line blanketed
SEDs. This calculation is always performed in LTE even when synthesizing the 
continuum of a NLTE model because PHOENIX cannot omit the NLTE b-b transitions in 
the calculations when synthesizing a spectrum. In all spectral synthesis cases, 
the SEDs are sampled from $\lambda$ = 3000 to 13000 $\textrm{\AA}$ 
as shown in Table \ref{tab:Wavelength-grid-spacing}, with variable spacings,
$\Delta\lambda$, to approximately preserve the spectral resolution,
$R=\lambda/\Delta\lambda$, across the full range at $R\approx$ 300000
to 400000. In addition to this, the NLTE SEDs are sampled at 
supplementary points that ensure each NLTE spectral line 
is critically sampled; these points are automatically
distributed over each line by PHOENIX.

\paragraph{}

Finally, to increase the apparent formal numerical resolution of the
grid's $T_{\mathrm{eff}}$ range sampling, additional SEDs were linearly
interpolated between neighboring SEDs in the grid to reach a final
apparent temperature resolution of 25 K. This was done to smooth the
final fitted $T_{\mathrm{eff}}$ value versus degree of inhomogeneity relation.
However, we note that this does not decrease the formal uncertainty
of the $T_{\mathrm{eff}}$ determination. Linear interpolation
of the flux values was used instead of interpolating the log flux
values because at the small ratio of $\Delta T_{\mathrm{eff}}/T_{\mathrm{eff}}$,
the relative difference of the monochromatic flux ($\Delta F_{\lambda}/F_{\lambda}$)
between the two methods was less than 1.0 \% at all wavelength sampling
points.

\subsection{1.5D SED generation, post-processing, and analysis \label{sec:1.5D}}

Seventeen unique 1.5D SEDs were produced to serve as artificial 2D targets by 
linearly averaging two synthetic NLTE 1D SEDs. These were individually 
distinguished by the differences in the $T_{\mathrm{eff}}$ values of their warm 
and cool 1D components, $\Delta T_{\mathrm{1.5D}}$. This $\Delta T_{\mathrm{1.5D}}$ 
corresponds to the contrast of inhomogeneities. It was enforced that, 
for each 1.5D SED, the linear average of their component $T_{\mathrm{eff}}$ values 
be 4250 K, effectively giving each of the 1.5D SEDs a $T_{\mathrm{eff,R-J}}$ = 
4250 K. The production of 1.5D SEDs by the linear averaging of the component 
fluxes in this way results in continuum levels in the Rayleigh-Jeans tails being 
identical within a few percent, and was chosen to maintain a consistent 
observational property among all of the 1.5D SEDs.

\paragraph{}

Two different weighting schemes were used in creating the 1.5D SEDs: 1) Evenly 
weighting both components as a simple reference case; and 2) Weighting the hot component at 
a ratio of 2:1 to the cool component. This ratio approximately represents the relative solar 
surface area covered by granules and intergranular lanes respectively \citep{sheminova12}. 
These two methods simulate a variation in surface coverage, or filling factor (FF), 
of the hot and cool features. Hereafter, the evenly weighted method is referred to 
as having a 1:1 FF, and the method with the hot component being weighted at 2:1 
as referred to as having a 2:1 FF. To fully explore the effects of horizontal 
inhomogeneity, we take the upper limit of 300 K for the RMS temperature difference 
across the features found from 3D modeling, and double it to 600 K, for a maximum 
realistic value of $\Delta T_{\mathrm{1.5D}}$.  We then extend this value up to 
1000 and 1050 K, for the 1:1 and 2:1 FF respectively, to test how large the RMS 
temperature variations would have to be before 1D estimates of $T_{\mathrm{eff}}$ 
become very misleading, with significant departures from $T_{\mathrm{eff,S-B}}$. 
Table \ref{tab:1.5D-SED} displays the components 
used for the two schemes, at each $\Delta T_{\mathrm{1.5D}}$ that exactly 
preserves both the $T_{\mathrm{eff,R-J}}$ value as 4250 K and the respective FF, 
as well as displaying the resultant 1.5D $T_{\mathrm{eff,S-B}}$ 
as found from the average of the components' bolometric luminosities.

\paragraph{}

\begin{figure}
\plotone{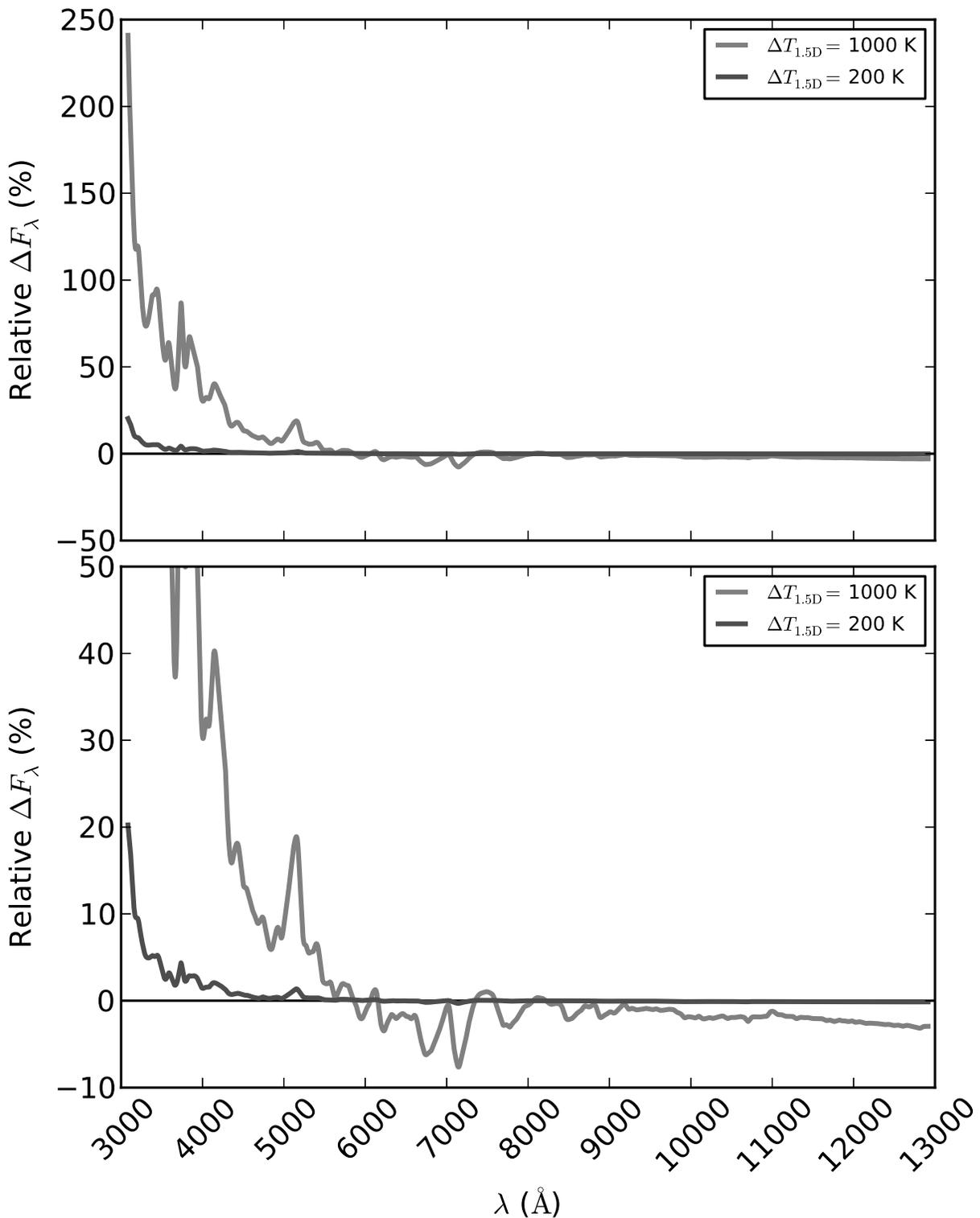}
\caption{Relative difference between two FF 1:1 1.5D spectra, $\Delta T_{\mathrm{1.5D}}$ = 200 and 1000 K, and 1D spectra with equivalent $T_{\mathrm{eff,S-B}}$ = 4254 and 4336 K, respectively. The bottom panel shows an expanded y-scale view to reveal additional detail not visible on the larger scale of the top panel.\label{fig:safe-zone}}
\end{figure}

We attempted to recover the $T_{\mathrm{eff}}$ values of the 1.5D
stars using three different methods of fitting 1D models: UBVRI photometry,
spectrophotometry, and spectroscopy. Common to each method of post-processing,
the 1.5D SEDs were interpolated from their initial wavelength distribution
to a new distribution with a constant $\Delta\lambda$ of 0.006 $\textrm{\AA}$.
As a preliminary test to see which diagnostics would be expected to return 
$T_{\mathrm{eff}}$ values close to $T_{\mathrm{eff,S-B}}$, and which would return 
values differing from $T_{\mathrm{eff,S-B}}$, we examined the relative difference 
between two of our 1:1 FF spectra, $\Delta T_{\mathrm{1.5D}}$ = 200 and 1000 K, 
and 1D models with their equivalent $T_{\mathrm{eff,S-B}}$. These spectra were 
convolved with a Gaussian kernel of  FWHM = 50 $\textrm{\AA}$, representative of 
the nominal resolution element of the observed spectrophotometry in the Burnashev 
catalogue \citet{burnashev85}, prior to taking the differences. While there is 
little difference between the 1D and 1.5D in the near-IR and parts of the visible wavebands, 
they begin to diverge rapidly in the UV as seen in Fig. \ref{fig:safe-zone}. This suggests 
that diagnostic tools focused on redder wavelengths are more likely to return 
$T_{\mathrm{eff}}$ values close to $T_{\mathrm{eff,S-B}}$. A more detailed high-resolutions 
analysis at the level of individual spectral lines could reveal specific diagnostics 
that are robust against 2D effects, but was beyond the scope of this study.

\subsubsection{UBVRI photometry}

For each of the NLTE and LTE 1D and NLTE 1.5D SEDs, photometric colors were
produced using Bessel's updated Johnson-Cousins UBVRI photometry \citep{bessell90}. 
The transmission data for each filter were interpolated to the wavelength
distribution using quadratic splines. Five different color
indices, U$_{x}$-B$_{x}$, B-V, V-R, V-I, and R-I, were calculated
to milli-magnitude precision for each of the SEDs. It should be noted
that the U band is of limited accuracy as the short wavelength cutoff
in the observational U band is not defined by the instrumental configuration,
but rather ozone absorption in Earth's atmosphere. Because ozone column
density varies by $\sim$20 \% with geographic location and season,
it is difficult to accurately reproduce U band photometry synthetically
from models. These values were calibrated with a PHOENIX NLTE synthesized
SED for the standard star Vega ($\alpha$ Lyr, HR7001, HD172167),
using a single-point photometric calibration independent of color.

\paragraph{}

\begin{figure}
\plotone{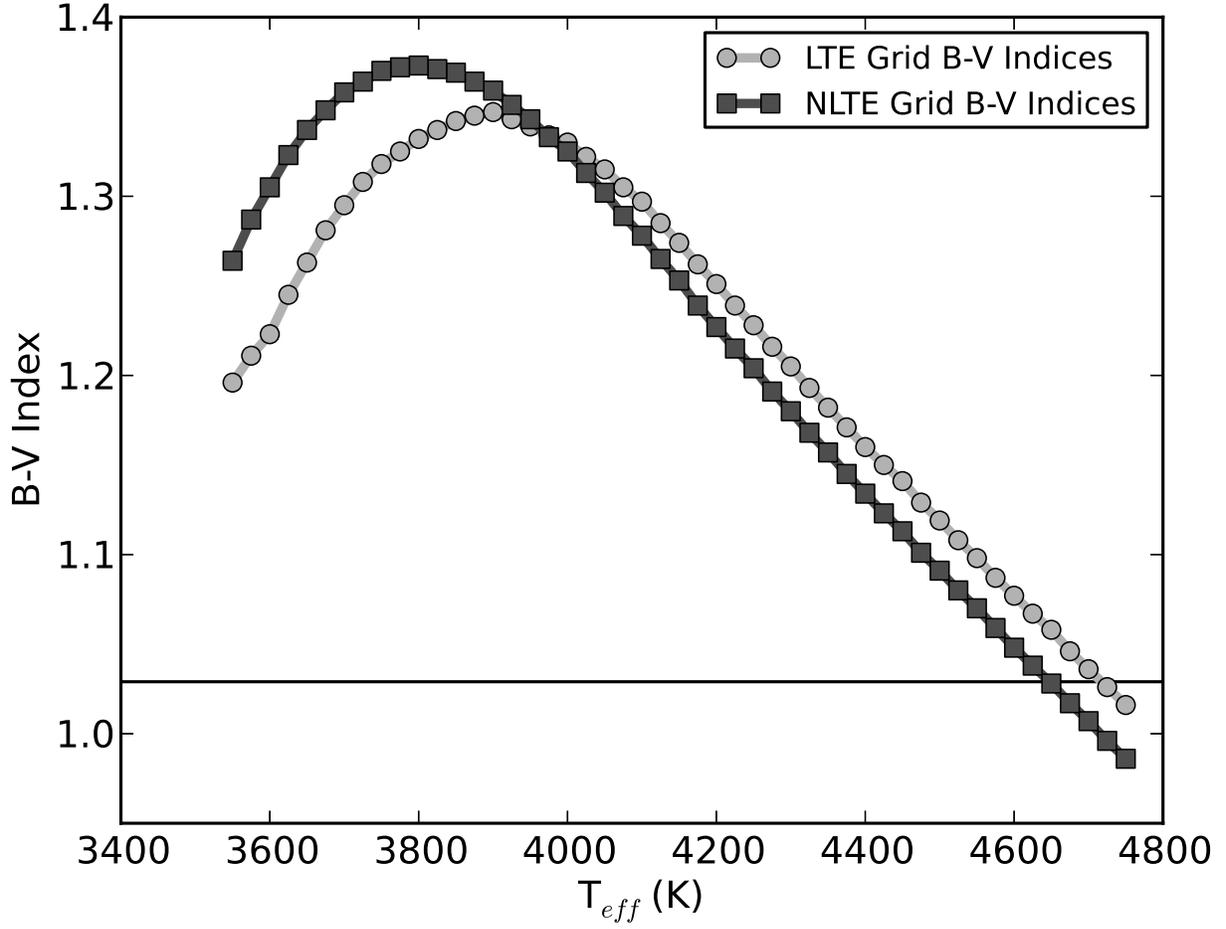}
\caption{B-V color index values for the LTE and NLTE 1D SEDs. The turnover occurs at $T_{\mathrm{eff}}$ = 3900 K for the LTE SEDs and at $T_{\mathrm{eff}}$ = 3800 K for the NLTE SEDs. The solid black line indicates the value of the B-V index for the $\Delta T_{\mathrm{1.5D}}$ = 1000 K, 1:1 FF 1.5D SED.\label{fig:turnover}}
\end{figure}

While three of the five color indices for the 1D SEDs behaved as
expected from the behavior of $B{}_{\lambda}$, with their values
monotonically increasing for decreasing $T_{\mathrm{eff}}$ values over the
range of the 1D SEDs, both the U$_{x}$-B$_{x}$ and B-V indices stopped
increasing and started decreasing at $T_{\mathrm{eff}}$ = 3900 K
for the LTE SEDs, and at $T_{\mathrm{eff}}$ = 3800 K for the NLTE
SEDs. This turnover is observed in stars, but between $T_{\mathrm{eff}}$
= 3540 K and 3380 K for the U-B index \citep{cox00}. The phenomenon
is caused by spectral features in the B filter growing in strength
more rapidly with decreasing $T_{\mathrm{eff}}$ values than those in the
U filter, and reverses the trend in the color index expected from
$B{}_{\lambda}$. Likewise, the spectral features in the V filter
grow more rapidly than those in the B filter. In this case, the incorrect
prediction of the $T_{\mathrm{eff}}$ value of the turnover is caused
by the excess blue and UV flux common to PHOENIX models of cool red
giant stars \citep{short03}. Fig. \ref{fig:turnover} shows the
B-V index value trend, including turnover, for both the LTE and NLTE
1D SEDs.

\paragraph{}

To quantify the errors from fitting 1D NLTE and LTE photometry to NLTE 2D stars, 
the library of 1D color index values was fitted to each of the 1.5D color index 
values. The closest matching $T_{\mathrm{eff}}$ value was found by means of 
inspection. Whichever 1D SED had the smallest difference between its value and 
the 1.5D value for each index was chosen as the match. For example, Fig. \ref
{fig:turnover} shows the value of the B-V index for the 1.5D SED of 1:1 FF with $
\Delta T_{\mathrm{1.5D}}$ = 1000 K including the region of closest match to 1D B-V 
values. Because of the degeneracy caused by the turnover in the U$_{x}$-B$_{x}$
and B-V indices and the limiting $T_{\mathrm{eff}}$ numerical resolution of the 1D 
grid, it was possible that the best matching $T_{\mathrm{eff}}$ value
would fall outside of the range of temperatures enclosed by the cool
and hot 1D components, and would not be the correct match. Automatically 
limiting the best match to within the components' $T_{\mathrm{eff}}$
range was determined to be too restrictive, so the closest three matches for 
both of the indices were found and ranked in order, whereupon the correct best 
match was chosen by inspection to logically fit the $T_{\mathrm{eff}}$ range.

\subsubsection{SEDs and Spectra}

\paragraph{Spectrophotometry}

\begin{figure}
\plotone{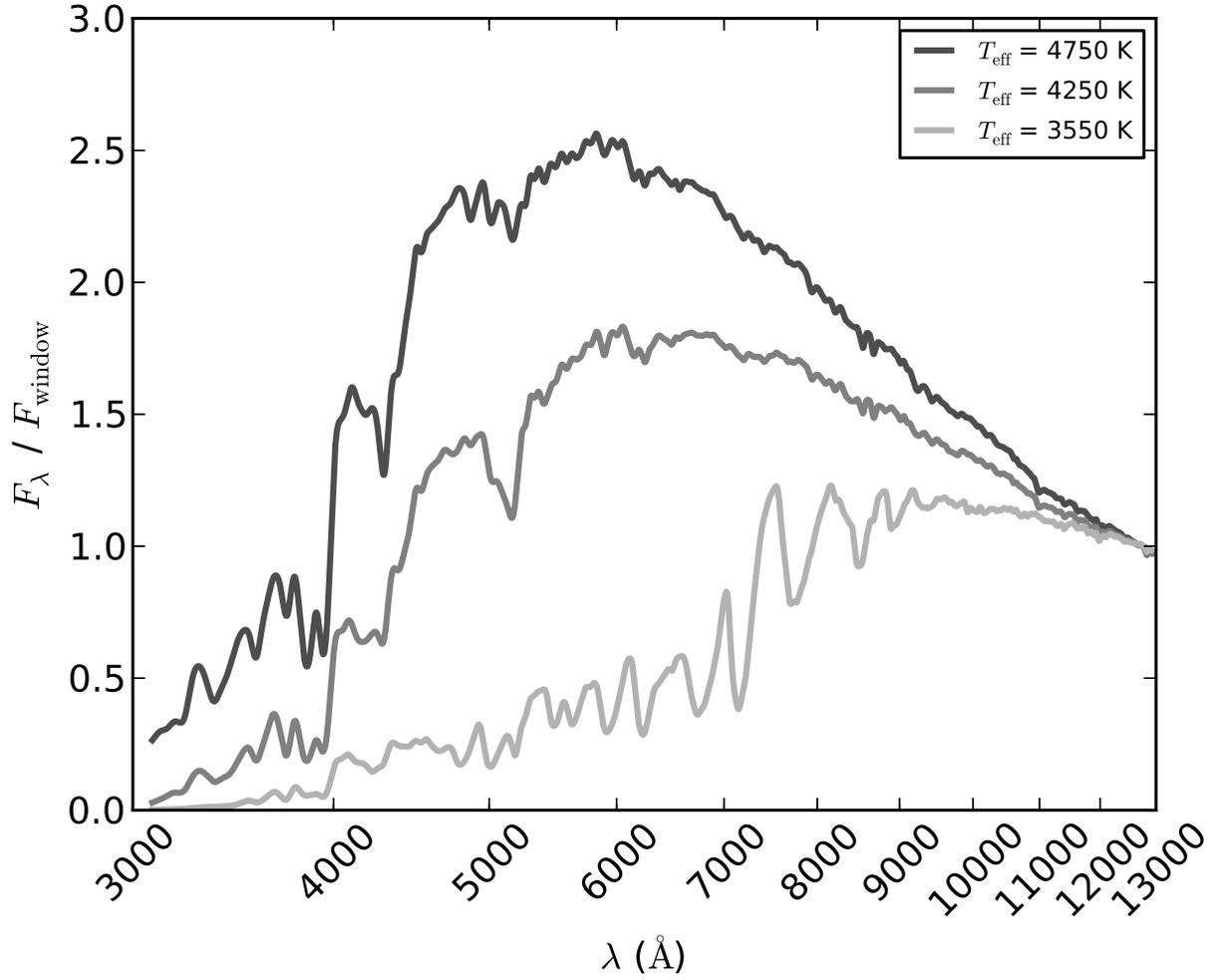}
\caption{Relative version of three NLTE SEDs with $T_{\mathrm{eff}}$ values of 4750 K, 4250 K, and 3550 K.\label{fig:pinned}}
\end{figure}

Absolute surface flux SEDs, $F_\lambda$, of the 1.5D models were fitted with 1D 
SEDs to find the closest match to the the total energy radiated by the 1.5D 
stars; the relative flux SEDs were fitted with 1D relative SEDs to find the 
closest match to the overall shape of the 1.5D SEDs. For the relative flux SEDs,
the surface flux values were normalized to the average flux within a 10 
$\textrm{\AA}$ window ($F_{\mathrm{window}}$) located between $\lambda$ = 12750
and 12760 $\textrm{\AA}$ in the Rayleigh-Jeans tail of each SED. A 10 $\textrm{\AA}$ 
window was chosen over a single data point to avoid having selected a point
that unreliably represents the continuum level in different SEDs. This 
window was chosen over others in the Rayleigh-Jeans tail because, it regularly 
had the fewest and weakest spectral features contained within its bounds.
Fig. \ref{fig:pinned} displays the relative version of the NLTE 
SEDs with $T_{\mathrm{eff}}$ values of 4750 K, 4250 K, and 3550 K.

\paragraph{Spectroscopy}

\begin{figure}
\plotone{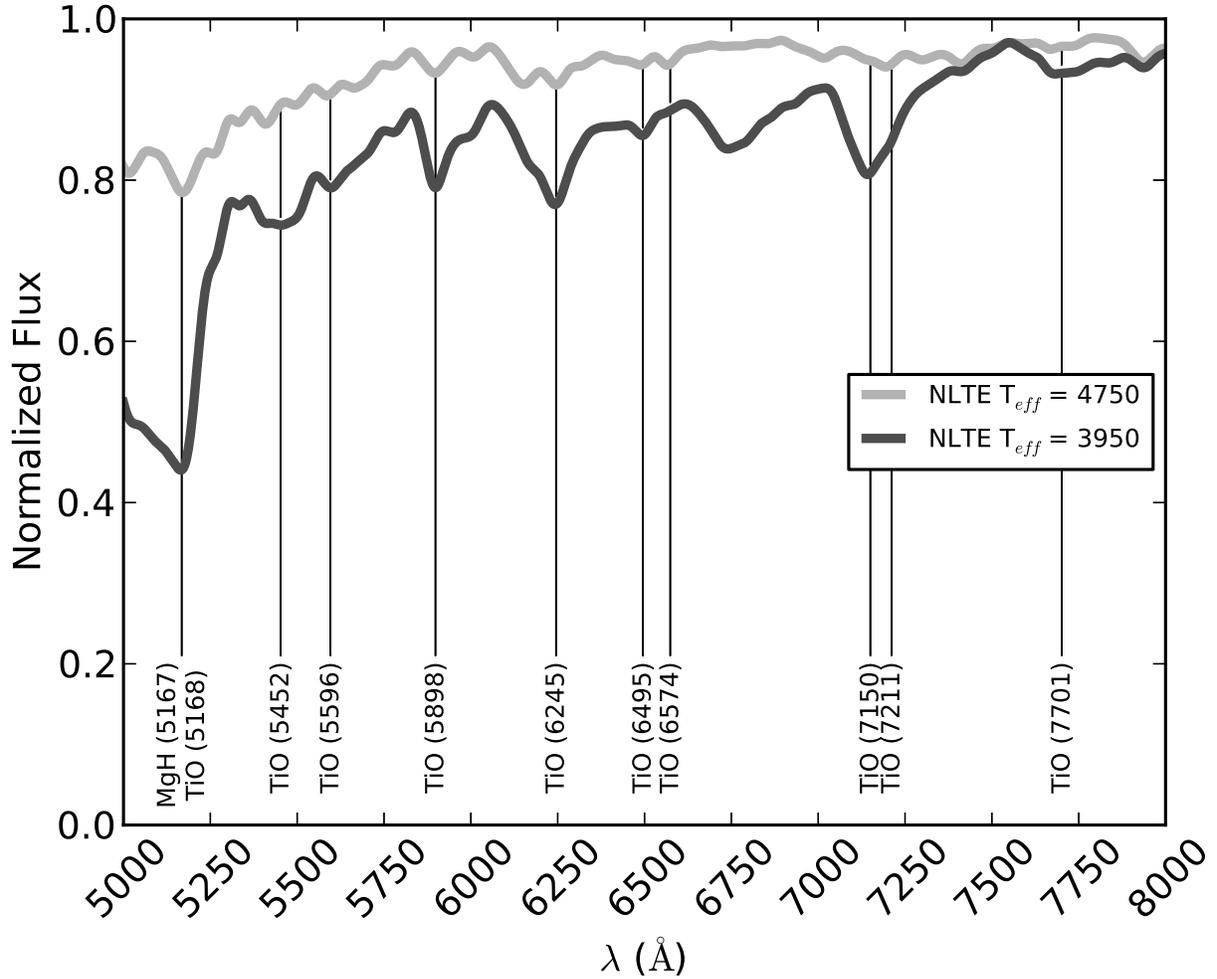}
\caption{Flattened NLTE 1D spectra with $T_{\mathrm{eff}}$ = 4750 K and 3950 K, restricted to a wavelength range of $\lambda$ = 5000 to 8000 $\textrm{\AA}$. Strong molecular absorption features in the region have been identified to illustrate the difference in temperature dependence of the MgH and TiO features.\label{fig:lineid}}
\end{figure}

Continuum normalized spectra were fitted with 1D spectra to determine the closest 
match based upon the relative strength of the spectral features. Continuum 
normalization was done for the 1D spectra by dividing the blanketed spectra by the 
corresponding continuum spectra. For the 1.5D SEDs, the same continuum normalization 
process was used, where the 1.5D synthetic continua were generated the same way that 
the 1.5D SEDs were, by linearly averaging the two corresponding 1D synthetic continua 
together. Initial fitting of the full wavelength range showed that the 
1.5D spectra could be fit well by 1D spectra in the blue/UV and the near IR wavebands, 
but showed a large discrepancy over most of the optical waveband between $\lambda$ = 
4800 and 8000 $\textrm{\AA}$. This discrepancy, caused primarily by the distinctive 
TiO molecular absorption bands present in cool star spectra \citep{davies13}, 
prompted a second fitting diagnostic for the continuum normalized spectra, by 
restricting the fit to the wavelength range between $\lambda$ = 5500 and 8000 
$\textrm{\AA}$. An MgH band centered at 5167 $\textrm{\AA}$ was found to overlap 
with the TiO band near this wavelength range, and depended differently on $T_{\mathrm{eff}}$
than the TiO bands, affecting the quality of the fits. The overlapping
MgH and TiO bands and additional nearby TiO bands are displayed in
Fig. \ref{fig:lineid}, illustrating the different temperature dependence
of the features.

\paragraph{}

Each of the high resolution 1D and 1.5D SEDs were convolved with a Gaussian 
kernel of  FWHM = 50 $\textrm{\AA}$. The smoothed SEDs were then re-sampled to
a much coarser $\Delta\lambda$ spacing to more accurately reflect
the number of degrees of freedom available when comparing with an
observed SED, but still fine enough to critically sample every feature
remaining after the smoothing convolution. This amounted to a sample
of 4000 wavelength points spaced between $\lambda$ = 3100
and 12900 $\textrm{\AA}$, with 100 $\textrm{\AA}$ at each end of
the original wavelength range having been lost to convolution edge
effects. For both $F_\lambda$ and relative surface fluxes, the new points were
evenly distributed in logarithmic space instead of linear space, amounting
to taking the log of the upper and lower bounds of the wavelength
region, distributing the new sample points evenly between these values,
and then exponentiating everything to return to linear space. This
resulted in a distribution of wavelength points with smaller $\Delta\lambda$
at shorter wavelengths and larger $\Delta\lambda$ at longer wavelengths.
The motivation behind this was to attribute additional weight to spectral
features located in the blue end of the SED when determining a best
fit, without arbitrarily weighting select wavelength regions more
heavily or determining a best fit for these regions separate from
the rest of the SED altogether. Extra weighting was attributed to
the blue band of the wavelength range because there are more spectral
features per $\Delta\lambda$ interval there, and it is more sensitive
to changes in temperature than the red band, making it a more sensitive
diagnostic tool for determining $T_{\mathrm{eff}}$. The continuum normalized
spectra were re-sampled in linear wavelength space to not grant any of the 
spectral features additional weight in the final fitting process. Unlike the 
spectrophotometric SEDs, the relative strengths of individual spectral features 
are of interest here, not the overall shape of the spectrum.

\paragraph{}

In all cases, the best fit 1D spectrum was determined by minimizing
a modified Pearson $\chi^{2}$ test statistic, of the form
\begin{equation}
\chi^{2}=\frac{1}{n}\sum_{i=1}^n\frac{\left(1-M_{i}/O_{i}\right)^{2}}{M_{i}/O_{i}}.\label{eq:Chi sq}
\end{equation}
where $n$ is the number of degrees of freedom, $O_{i}$
is the observed frequency of a phenomenon to be fitted, and $M_{i}$
is the modeled or expected frequency of the phenomenon. The 1D and
1.5D fluxes were treated as the modeled and observed frequencies, respectively,
because a spectral flux is in some sense the frequency at which photons
of given energies emerge from stars. For each 1.5D SED, every 1D SED was 
individually treated as the model value to create a test statistic value. The 
1D SED with the minimum $\chi^{2}$ value was chosen as the best fitting SED from
which the 1.5D $T_{\mathrm{eff}}$ value was inferred. In each fitting case,
the result was determined to be significant at a confidence level
of $p$ = 0.05. As a check for self consistency, the 1D NLTE fitted results are 
expected to be bounded by the hot and cold components' $T_{\mathrm{eff}}$ 
values for a given 1.5D star, but this restriction is not expected of the 1D 
LTE fitted results.

\section{RESULTS OF 1D $T_{\mathrm{eff}}$ FITTING \label{Results}}

For all cases except one, the best fitting $T_{\mathrm{eff}}$
values for the 1.5D stars showed an increasing trend with increasing
$\Delta T_{\mathrm{1.5D}}$. This is expected from the non-linear
dependence of $B_{\lambda}$ on temperature, where a hotter $T_{\mathrm{eff}}$
value produces disproportionately more flux than a cooler one, and
the averaged flux in 1.5D stars will be more than a $T_{\mathrm{eff}}$
value of 4250 K would suggest for a 1D star. This effect is even more
pronounced at bluer wavelengths. Taking the derivative of $B_{\lambda}$
with respect to temperature, 
\begin{equation}
\frac{\partial B_{\lambda}}{\partial T}=\frac{2h^{2}c^{3}}{\lambda^{6}k_{B}T_{\mathrm{eff}}^{2}}\frac{e^{hc/\lambda k_{B}T}}{\left(e^{hc/\lambda k_{B}T}-1\right)^{2}},\label{eq:dB/dT}
\end{equation}
inspection reveals that for changing temperatures, $B_{\lambda}$
exhibits larger relative changes at shorter wavelengths, causing the
flux at the blue end of the spectrum to increase disproportionately
faster than the rest of the wavelength range with increasing $T_{\mathrm{eff}}$.

\paragraph{}

For the case of the TiO bands fitted in the continuum normalized spectra, the
reverse trend was instead seen; a decreasing best fit $T_{\mathrm{eff}}$ value
with increasing $\Delta T_{\mathrm{1.5D}}$. This is expected from
the non-linear dependence of molecule formation on gas temperature
\citep{uitenbroek11}. All fitting results are considered as having
a formal uncertainty of $\delta T=\pm$25 K, one half of the original
numerical temperature resolution of the 1D grid. The fit results are summarized in 
Tables \ref{tab:Master-Results-Table-1} and \ref{tab:Master-Results-Table-2}, 
while Tables \ref{tab:Master-Difference-Table-1} and 
\ref{tab:Master-Difference-Table-2} display the differences between the fit 
results and the 1.5D $T_{\mathrm{eff,S-B}}$. Results for comparing with 1D LTE 
spectra are limited to diagnostics only involving V-band photometry and redder, 
or equivalent wavelength ranges, as discussed in section \ref{PHOENIX}.

\subsection{UBVRI photometry \label{sec:UBVRI-photometry}}

\begin{figure}
\plotone{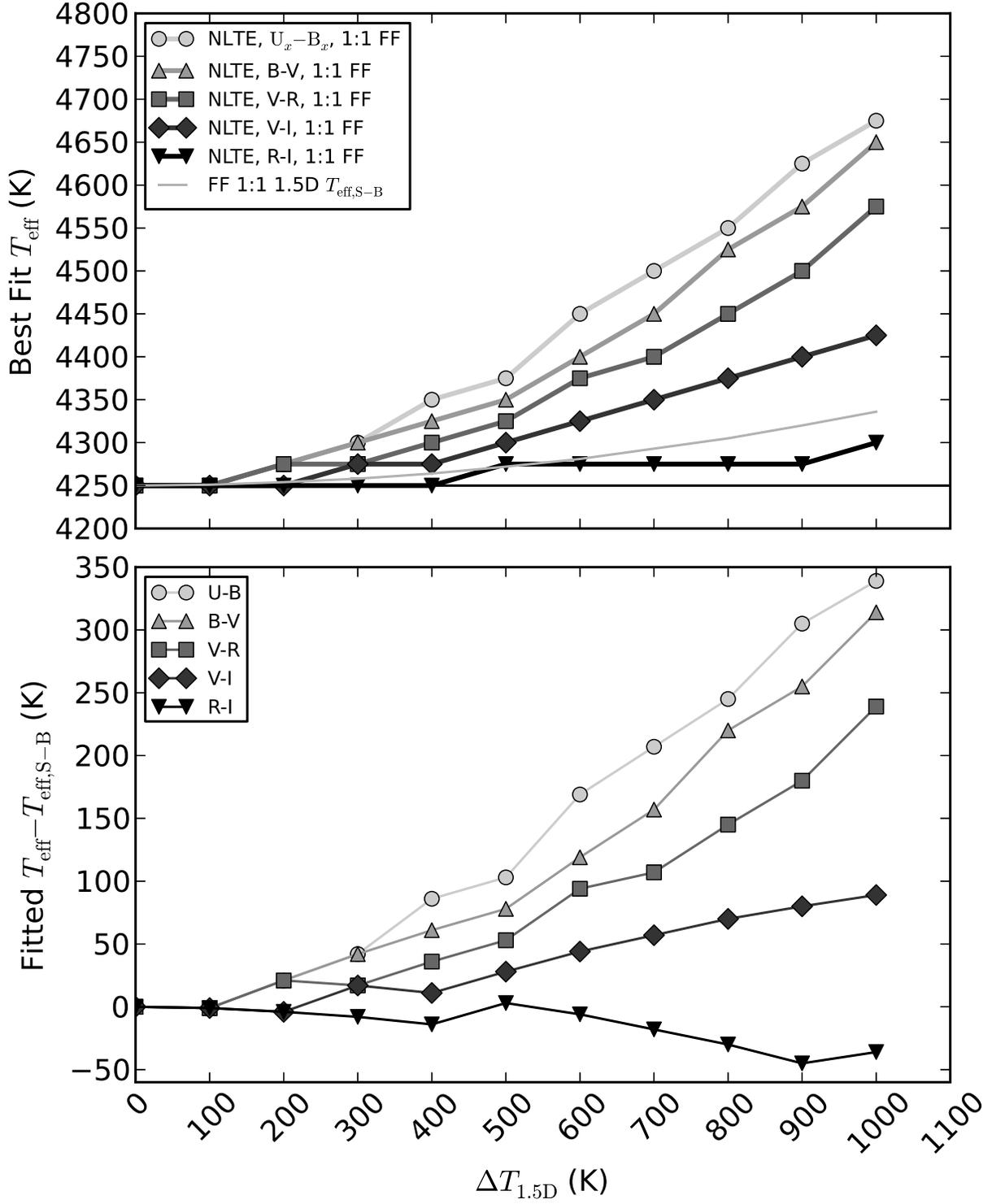}
\caption{Inferred $T_{\mathrm{eff}}$ values as functions of the difference in 1.5D component $T_{\mathrm{eff}}$, $\Delta T_{\mathrm{1.5D}}$, for five photometric color indices and 1:1 FF. The top image displays the fit values and the 1.5D $T_{\mathrm{eff,S-B}}$, with the solid black line representing $T_{\mathrm{eff,R-J}}$ = 4250 K, and the bottom displays these difference between these values and $T_{\mathrm{eff,S-B}}$ for a given $\Delta T_{\mathrm{1.5D}}$.\label{fig:NLTE-Photometry-Results}}
\end{figure}

\begin{figure}
\plotone{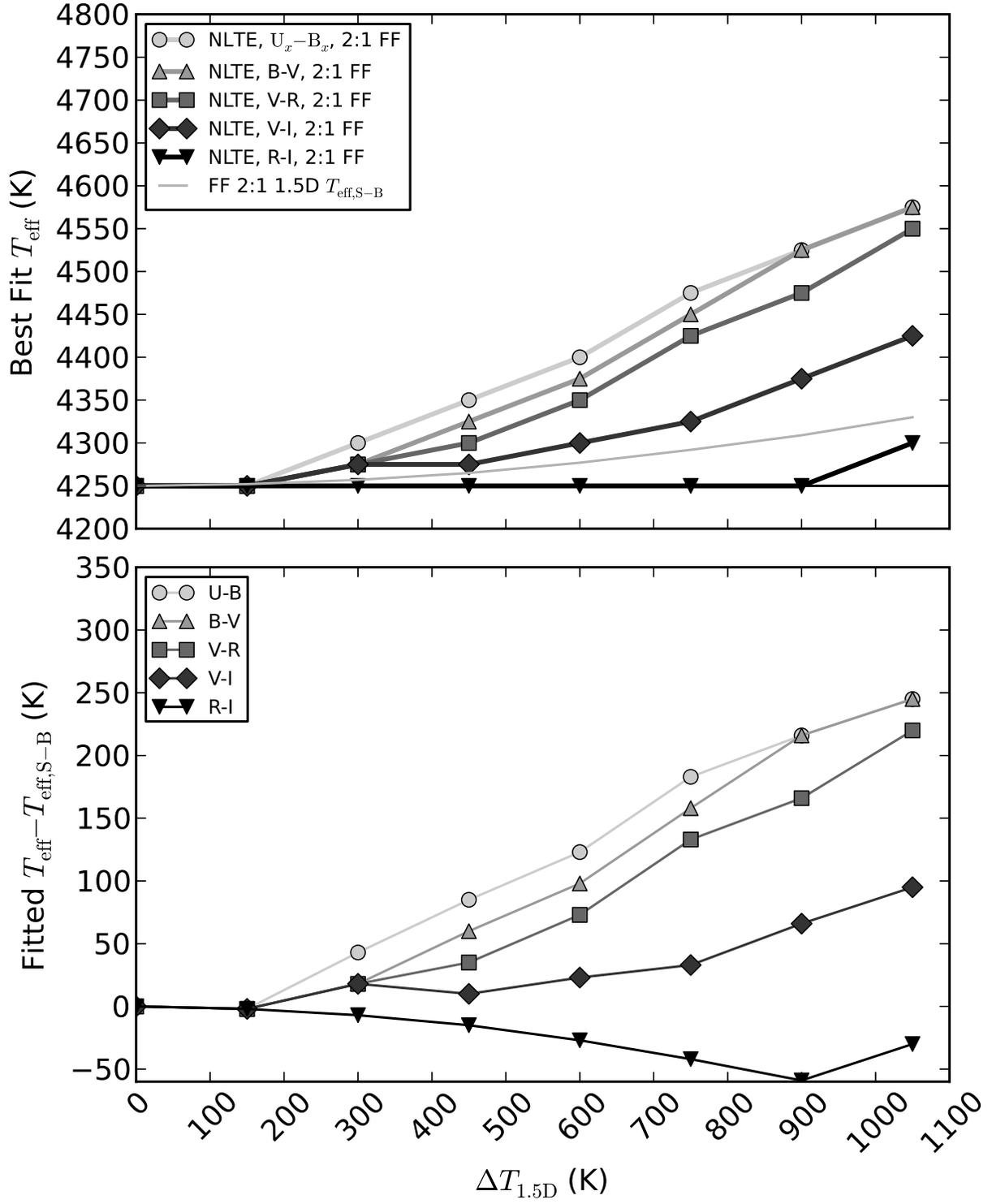}
\caption{Same as Fig. \ref{fig:NLTE-Photometry-Results}, but for 2:1 FF.\label{fig:NLTE-Photometry-Results-2:1}}
\end{figure}

Figs. \ref{fig:NLTE-Photometry-Results} and \ref{fig:NLTE-Photometry-Results-2:1}
presents each of the five photometric color indices' NLTE best fitting $T_{\mathrm{eff}}$
values as functions of $\Delta T_{\mathrm{1.5D}}$ for all of the
1.5D stars. 
Each of the color indices shows increasing best fit $T_{\mathrm{eff}}$
values for the 1.5D stars with increasing $\Delta T_{\mathrm{1.5D}}$.
The slope of the $T_{\mathrm{eff}}\left(\Delta T_{\mathrm{1.5D}}\right)$
relation is different for the five indices, creating, for a given
1.5D star, a spread of best fit $T_{\mathrm{eff}}$ values among the
indices that grows with increasing $\Delta T_{\mathrm{1.5D}}$. The
slopes of the individual color indices' fits are steeper for the
bluer indices, and nearly flat for the R-I index over the $\Delta T_{\mathrm{1.5D}}$
range. Such a spread in the fits shows that horizontal inhomogeneity
effects are non-constant across the spectrum. This spread cannot be
resolved at a 25 K resolution for $\Delta T_{\mathrm{1.5D}}\leq$
150 K, and grows as large as 375 K for 1.5D stars with 1:1 FF, and
as large as 250 K for 2:1 FF.

\paragraph{}

For the 2:1 FF fits, the spread is not as large as the 1:1 FF, because,
while the 2:1 1.5D star is created with more hot material than cool
material, the $T_{\mathrm{eff}}$ values of both components are lower
than the respective components of a 1:1 1.5D star having the same
$\Delta T_{\mathrm{1.5D}}$, outweighing the contribution of more
hot material by having less flux to contribute from the hot component.
This resulted in either the same or lower temperature fits for a given
$\Delta T_{\mathrm{1.5D}}$. Additionally, because the cool component
of the 2:1 1.5D stars has a lower $T_{\mathrm{eff}}$ value than that
of a 1:1 1.5D star, the blue and UV regions are even more severely
line-blanketed, contributing even less flux than the difference in 1D 
$T_{\mathrm{eff}}$ values alone would suggest, and resulting in lower temperature
fits for the color indices at shorter wavelengths, noticeably U$_{x}$-B$_{x}$
and B-V.

\paragraph{NLTE}

\begin{figure}
\plotone{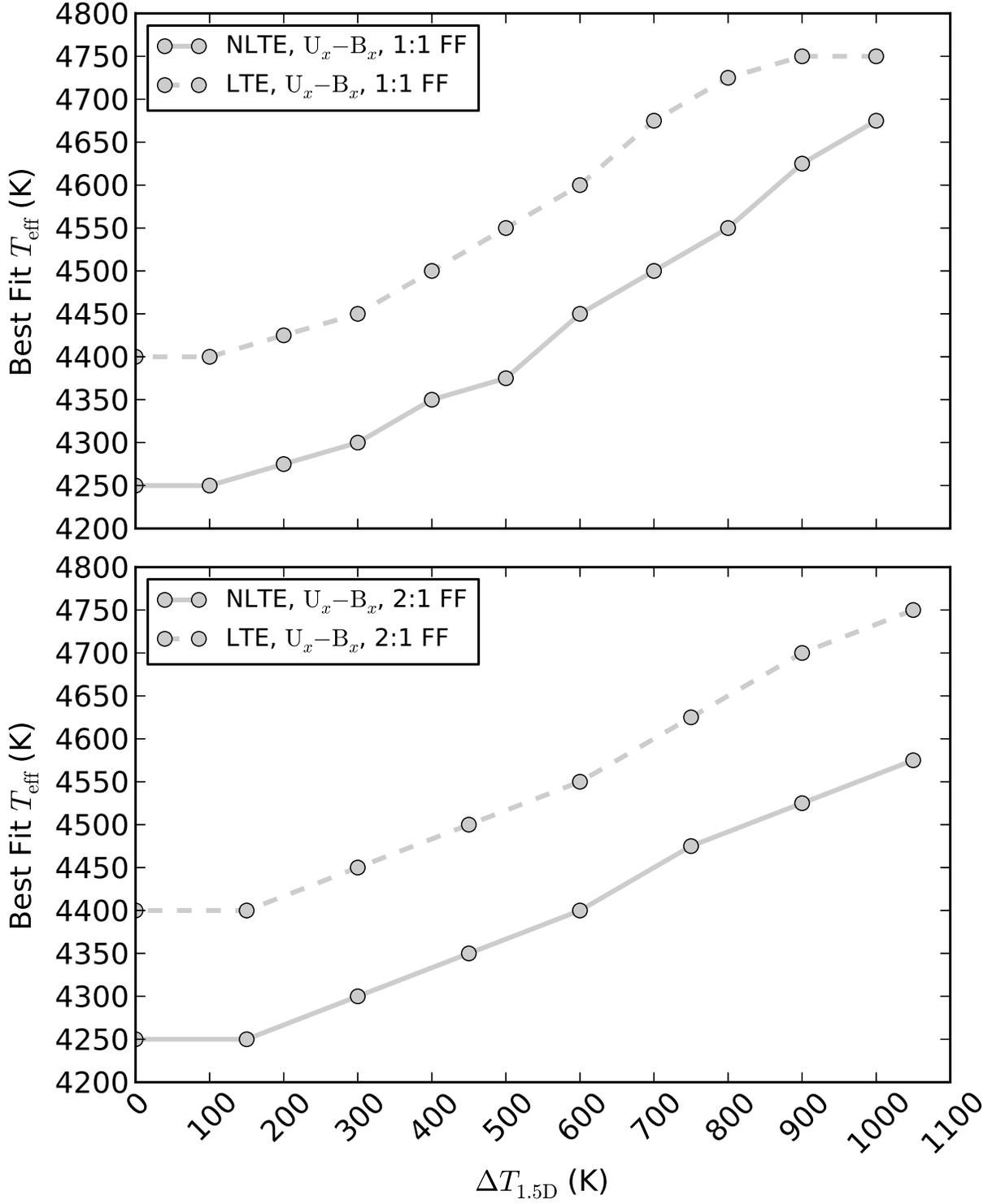}
\caption{U$_{x}$-B$_{x}$ index best fit $T_{\mathrm{eff}}$ value results. The top image contains results for 1.5D stars with 1:1 FF, while the bottom image contains those for 2:1 FF. Results from NLTE (solid lines) and LTE (dashed lines) 1D modeling. The data points for the LTE 1:1 FF fits at $\Delta T_{\mathrm{1.5D}}$ = 900K and 1000 K are at the upper limit of the 1D grid of models and may not accurately represent what the best fitting $T_{\mathrm{eff}}$ values may be.\label{fig:U-B-Results}}
\end{figure}

To quantify the magnitude of NLTE effects, we defined the quantity
$\Delta T_{\mathrm{NLTE}}$ as the difference between the fitted NLTE
$T_{\mathrm{eff}}$ value and the fitted LTE $T_{\mathrm{eff}}$ value
for a 1.5D star. Examining each color index individually for the
effects of NLTE, it is seen that both choices of FF return LTE best
fit $T_{\mathrm{eff}}$ values that are always hotter than the NLTE
best fits for a given $\Delta T_{\mathrm{1.5D}}$, and that $\Delta T_{\mathrm{NLTE}}$
is approximately constant within one numerical temperature resolution
unit for all choices of $\Delta T_{\mathrm{1.5D}}$ for each color
index. This constant value depends on color index, suggesting that
NLTE effects are non-constant across the spectrum, but they are temperature
independent over the range of $T_{\mathrm{eff}}$ values in the 1D
grid. For the U$_{x}$-B$_{x}$ index, the magnitude of $\Delta T_{\mathrm{NLTE}}$
is at least 100 K larger than the other indices because of the greater
flux in the blue and UV from NLTE Fe I overionization, forcing the
LTE SEDs to have a higher $T_{\mathrm{eff}}$ value to match the flux.
Fig. \ref{fig:U-B-Results} displays the results for the U$_{x}$-B$_{x}$
index, showing a nearly constant value for $\Delta T_{\mathrm{NLTE}}$
of -150 K to -175 K. The plateau in the LTE 1:1 FF 
$T_{\mathrm{eff}}\left(\Delta T_{\mathrm{1.5D}}\right)$
relation above $\Delta T_{\mathrm{1.5D}}$ = 800 K is not a breaking
of this constant $\Delta T_{\mathrm{NLTE}}$, but rather the best
fit $T_{\mathrm{eff}}$ values were restricted by the upper limit
of the 1D grid of models.

\subsection{Spectrophotometry}

\subsubsection{Absolute flux distributions}

Both choices of FF show an increasing trend in NLTE best fitting $T_{\mathrm{eff}}$
values for increasing $\Delta T_{\mathrm{1.5D}}$. For values of 
$\Delta T_{\mathrm{1.5D}}\leq$ 300 K, both choices of FF produce the same best 
fitting $T_{\mathrm{eff}}$ value for a given $\Delta T_{\mathrm{1.5D}}$ value, 
whereas for values of $\Delta T_{\mathrm{1.5D}}\geq$ 400 K, the 1:1 FF 1.5D SEDs 
produce hotter $T_{\mathrm{eff}}$ values than the 2:1 FF SEDs. The 1:1 FF
SEDs reached a best fitting $T_{\mathrm{eff}}$ value of 4450 K and
the 2:1 FF SEDs reached 4425 K at maximum $\Delta T_{\mathrm{1.5D}}$.

\paragraph{NLTE}

Comparing the LTE results with the NLTE results for both choices of FF, and 
for every value of $\Delta T_{\mathrm{1.5D}}$, the LTE best fitting $T_{\mathrm
{eff}}$ values are hotter than the NLTE values. Because NLTE stars produce more 
blue and UV flux for a given $T_{\mathrm{eff}}$ value than LTE stars, the LTE 1D 
SEDs returned higher best fitting $T_{\mathrm{eff}}$ values to match the
$F_\lambda$ value of the 1.5D SED. For the 1:1 FF 1.5D SEDs, $\Delta T_{\mathrm{NLTE}}$
varied between 25 K and 50 K, whereas for the 2:1 FF SEDs, $\Delta T_{\mathrm{NLTE}}$
was nearly constant at 50 K, only dropping to 25 K for $\Delta T_{\mathrm{1.5D}}$ =
900 K.

\subsubsection{Relative flux distributions}

\begin{figure}
\plotone{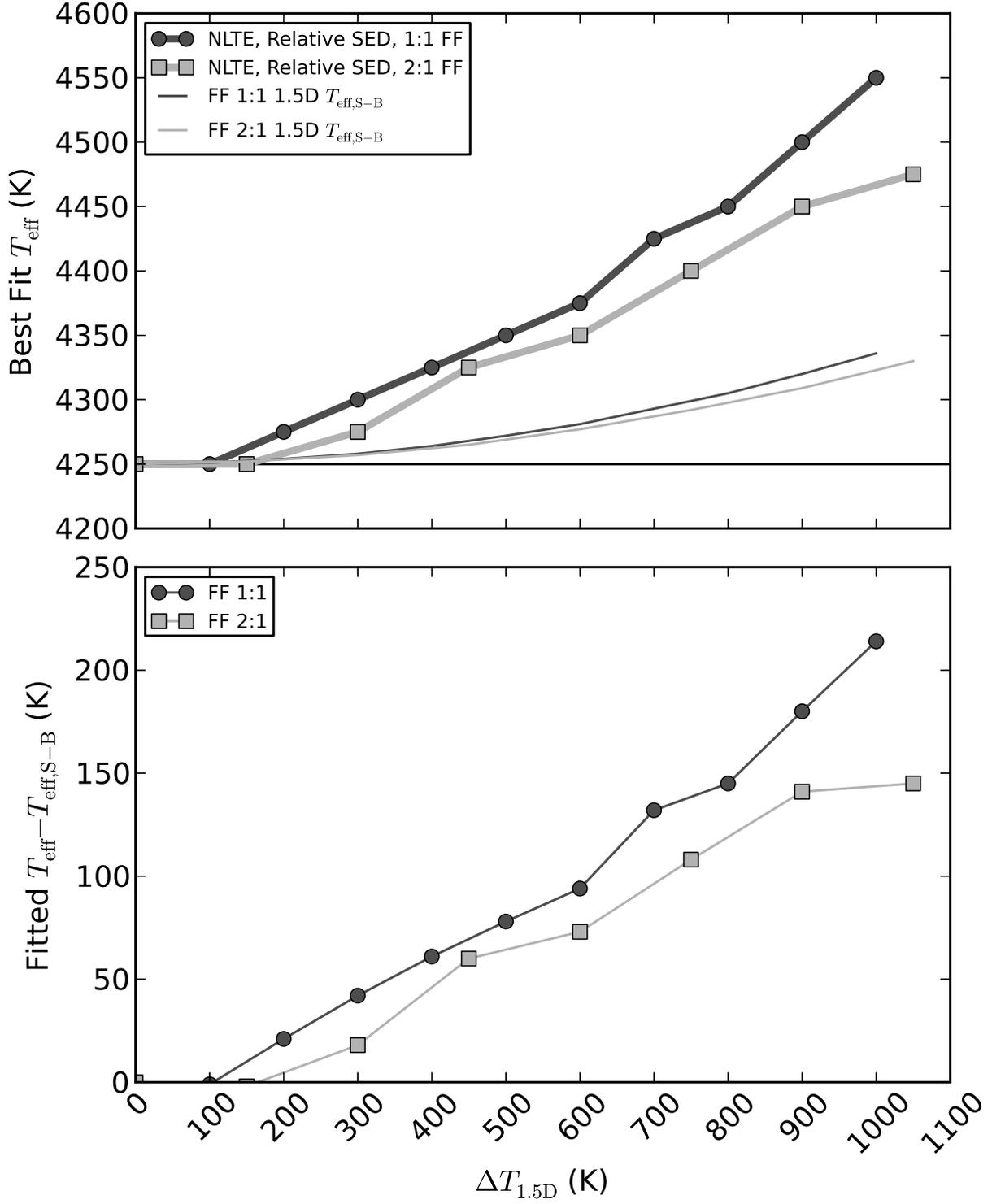}
\caption{Inferred $T_{\mathrm{eff}}$ from fitting relative surface flux SEDs with 1D NLTE SEDs. The top image displays the fit values and the 1.5D $T_{\mathrm{eff,S-B}}$, with the solid black line representing $T_{\mathrm{eff,R-J}}$ = 4250 K, and the bottom displays these difference between these values and $T_{\mathrm{eff,S-B}}$ for a given $\Delta T_{\mathrm{1.5D}}$.\label{fig:pin-NLTE-Results}}
\end{figure}

The results of fitting 1D NLTE relative flux SEDs to the 1.5D SEDs are presented in Fig. \ref{fig:pin-NLTE-Results}.
Both choices of FF show an increasing trend in best fitting $T_{\mathrm{eff}}$
values for increasing $\Delta T_{\mathrm{1.5D}}$ with the 1:1 FF
best fitting $T_{\mathrm{eff}}$ values being hotter than the 2:1
best fits at every value of $\Delta T_{\mathrm{1.5D}}>$ 100 K. The
1:1 FF SEDs reached a best fitting $T_{\mathrm{eff}}$ value of 4550 K and the 2:1
FF SEDs reached 4475 K at maximum $\Delta T_{\mathrm{1.5D}}$. 
Because the logarithmically spaced $\lambda$ grid was utilized to
place more weight on bluer wavelengths for the fits, the 1.5D SED
matches well with the best fitting 1D SED for $\lambda\lesssim$ 4500
$\textrm{\AA}$, but the 1D SED fails to predict the shape of the
1.5D SED around the peak of the spectrum; here, the 1.5D SED would
be fit by a SED with less flux at these wavelengths relative to $F_{\mathrm{window}}$,
suggesting a cooler $T_{\mathrm{eff}}$ value for the best fitting
1D SED. 

\paragraph{NLTE}

The LTE best fitting $T_{\mathrm{eff}}$ values are hotter than the
NLTE values for every value of $\Delta T_{\mathrm{1.5D}}$. For both
choices of FF, $\Delta T_{\mathrm{NLTE}}$ varied between 50 K and
75K.

\subsection{Spectroscopy}

\subsubsection{Continuum normalized spectra \label{sec:Continuum-normalized-spectra}}

The best fit $T_{\mathrm{eff}}$ values below $\Delta T_{\mathrm{1.5D}}$ =
500 K are approximately equal at each $\Delta T_{\mathrm{1.5D}}$
for both choices of FF, and only begin to differ above $\Delta T_{\mathrm{1.5D}}$ =
600 K, with the 1:1 FF 1.5D stars having hotter fitted values. The
inequality reaches a maximum $T_{\mathrm{eff}}$ value difference
of 100 K at the maximum $\Delta T_{\mathrm{1.5D}}$, with the 1:1
FF 1.5D stars reaching a maximum $T_{\mathrm{eff}}$ value of 4650
K. 

\paragraph{}

\begin{figure}
\plotone{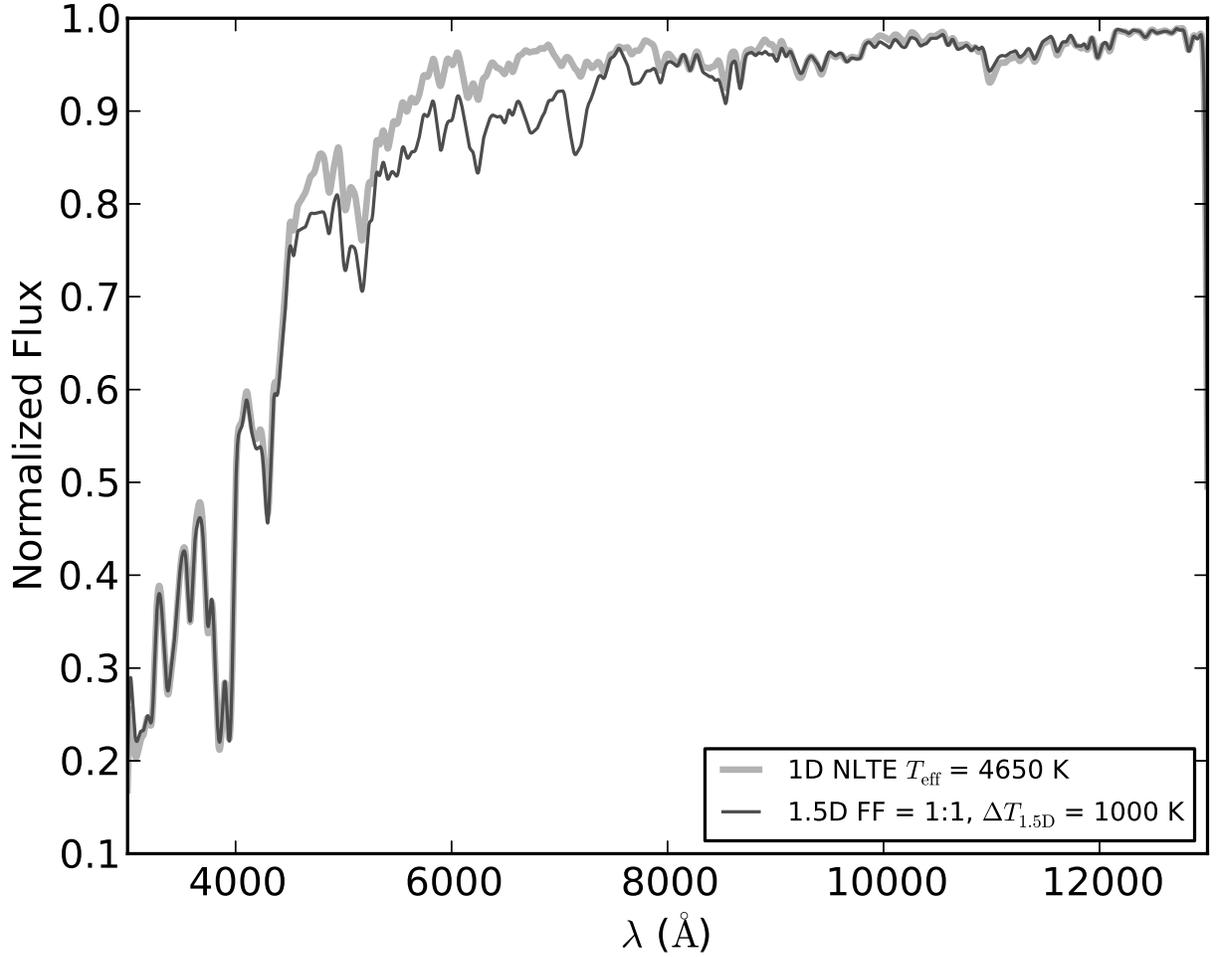}
\caption{Continuum normalized 1.5D 1:1 FF with the best fitting 1D NLTE spectrum. The drop off below $\sim$ 6000 $\textrm{\AA}$ is caused by the extreme level of line blanketing in the blue and UV. The poor fit between $\sim$ 4500 and 8000 $\textrm{\AA}$ is attributed to prominent TiO absorption features present in the cool component of the 1.5D spectrum.\label{fig:Rectified-comp}}
\end{figure}

Fig. \ref{fig:Rectified-comp} shows the 1.5D 1:1 FF continuum normalized spectrum
with $\Delta T_{\mathrm{1.5D}}$ = 1000 K plotted with the best fitting
1D NLTE spectrum and two bracketing 1D NLTE spectra. 
The best fitting 1D spectrum is seen to be a good fit to the 1.5D
spectrum for the blue and red ends of the $\lambda$ range, but is
a poor match between $\lambda\approx$ 4500 and 8000
$\textrm{\AA}$. In this region, the 1.5D spectrum exhibits stronger
spectral features than the best fitting 1D spectrum would imply, suggesting
that there is a cooler best fitting $T_{\mathrm{eff}}$ value for
this region. Additional fits omitting the poorly matched region were performed 
and the resultant $T_{\mathrm{eff}}$ values were found to differ by less than one 
numerical temperature resolution unit from the fits performed over the entire 
range, at maximum $\Delta T_{\mathrm{1.5D}}$.

\paragraph{NLTE}

Comparing the NLTE results with LTE results, it is seen that for both
FF values, the LTE best fit $T_{\mathrm{eff}}$ values are hotter
than the NLTE fits, but $\Delta T_{\mathrm{NLTE}}$ increases with
increasing $\Delta T_{\mathrm{1.5D}}$. For the 2:1 FF 1.5D stars, $\Delta T_
{\mathrm{NLTE}}$ reaches a maximum of 175 K, and for the 1:1 FF stars, it is 
potentially even larger but the LTE results plateau above $\Delta T_{\mathrm{1.5D}}$ 
= 800 K, similar to the U$_{x}$-B$_{x}$ photometric color index in
Section \ref{sec:UBVRI-photometry}, and for the same reasons.

\subsubsection{TiO bands}

\begin{figure}
\plotone{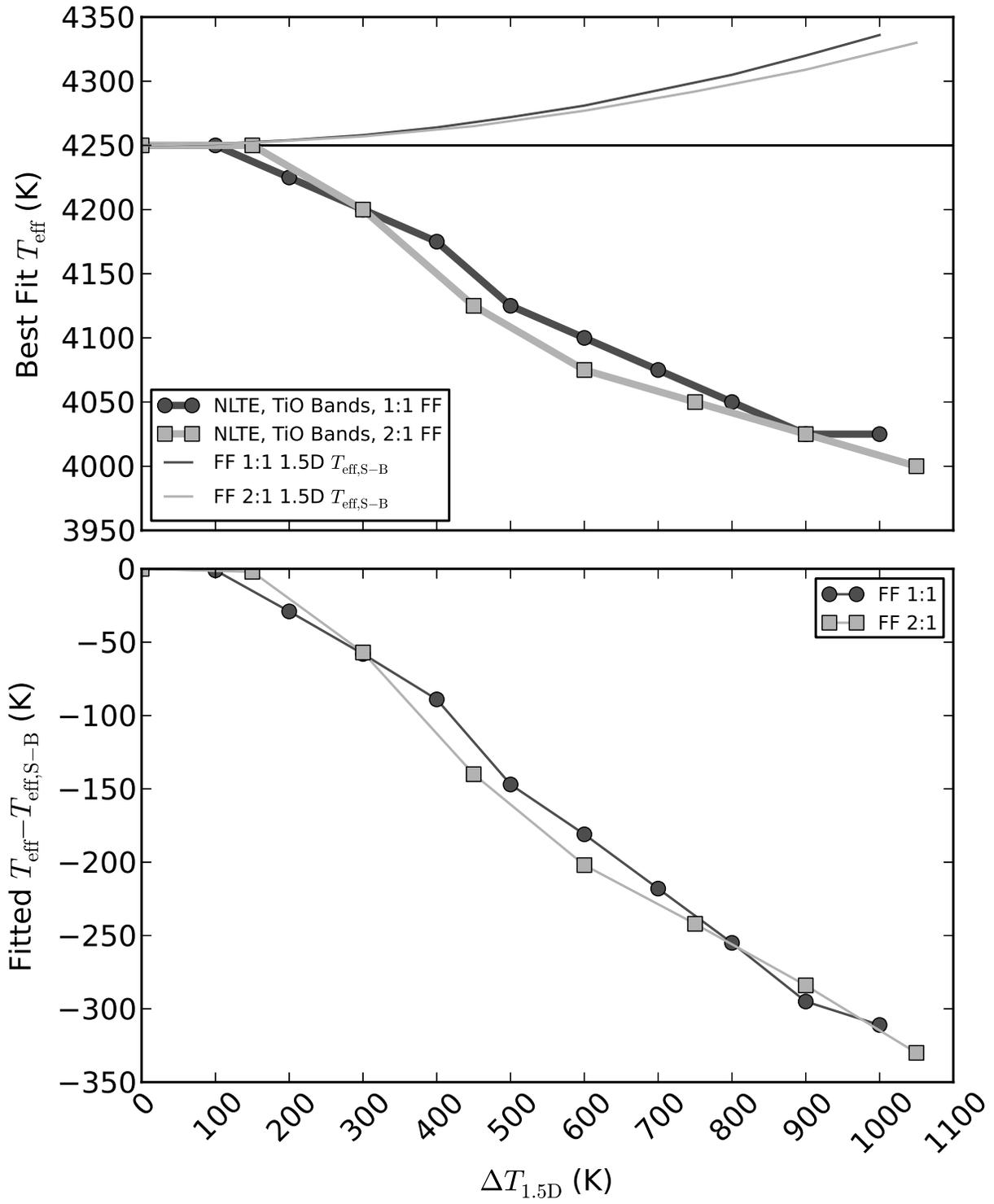}
\caption{Same as Fig. \ref{fig:pin-NLTE-Results} but for TiO bands in continuum normalized spectra.\label{fig:TiO-NLTE-Results}}
\end{figure}

\begin{figure}
\plotone{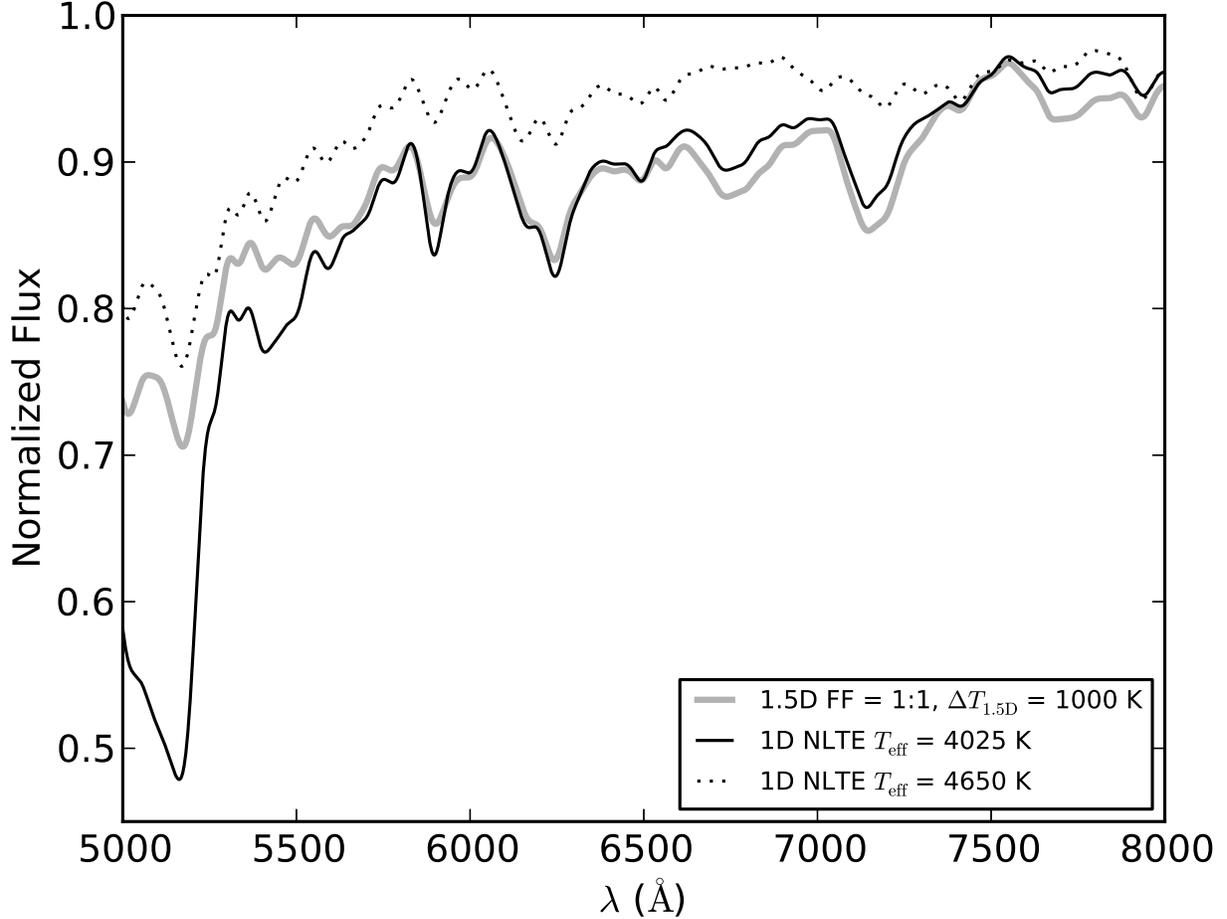}
\caption{Continuum normalized 1.5D 1:1 FF with the best fitting 1D NLTE spectrum (solid black) for the TiO band $\lambda$ range from 5500 to 8000 $\textrm{\AA}$. The best fitting 1D NLTE spectrum (dotted) from fitting the full $\lambda$ range is also shown to obviate the necessity of fitting this region independently.\label{fig:TiO-comp}}
\end{figure}

Fig. \ref{fig:TiO-NLTE-Results} shows NLTE best fitting $T_{\mathrm{eff}}$
values inferred from fitting to continuum normalized spectra restricted to the
TiO bands located between $\lambda$ = 5500 and 8000
$\textrm{\AA}$ for all 1.5D stars. A decreasing trend of fitted $T_{\mathrm{eff}}$ value
with increasing $\Delta T_{\mathrm{1.5D}}$ is observed, dropping as low as 
$T_{\mathrm{eff}}$ = 4000 K at maximum $\Delta T_{\mathrm{1.5D}}$. This agrees with 
the observation from section \ref{sec:Continuum-normalized-spectra} that the best fitting 
$T_{\mathrm{eff}}$ values for this $\lambda$ should be cooler than those found for
the same star when fitting the entire visible band. Fig \ref{fig:TiO-comp} 
shows the TiO fitting region of the 1.5D 1:1 $\Delta T_{\mathrm{1.5D}}$ = 1000 K 
with the best fitting 1D spectrum found for this region, as well as the 1D spectrum 
found for fitting the entire $\lambda$ range as a comparison. Both of the FF values
produce similar results when fit with NLTE 1D stars, the fits being
separated by no more than one numerical temperature resolution unit
at any $\Delta T_{\mathrm{1.5D}}$. The 1:1 FF 1.5D stars have the
hotter fitted $T_{\mathrm{eff}}$ values at $\Delta T_{\mathrm{1.5D}}$
values greater than 600 K. The TiO molecular bands grow in strength
so rapidly with decreasing $T_{\mathrm{eff}}$ that the hot component
does not dominate the fit in the TiO band region as it does for the
overall continuum normalized spectra. The cool component now dominates the shape,
with the hot component only mitigating the effect, pulling the fits
to lower temperatures with increasing $\Delta T_{\mathrm{1.5D}}$.

\paragraph{NLTE}

\begin{figure}
\plotone{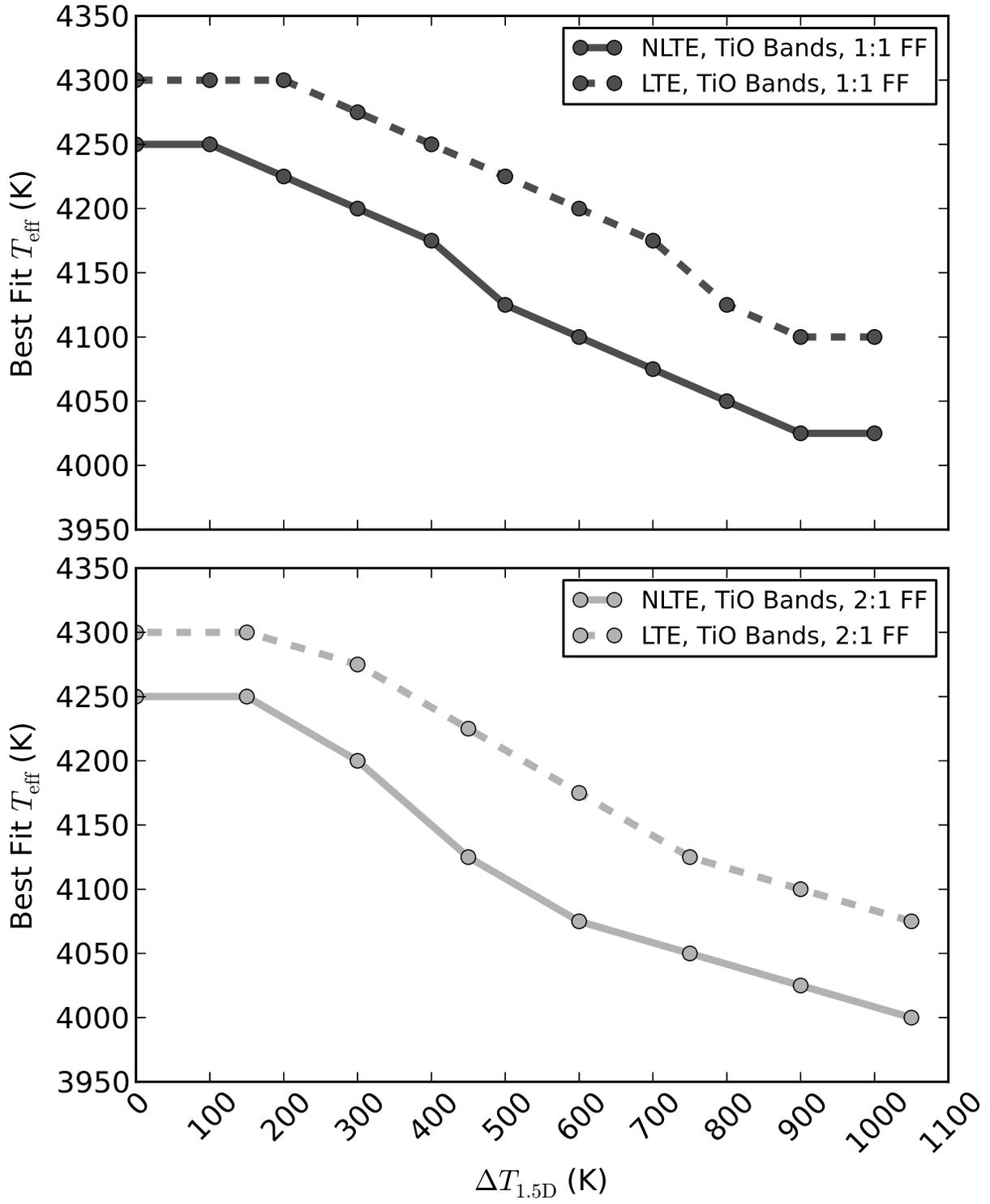}
\caption{Same as Fig. \ref{fig:U-B-Results} but for best fitted $T_{\mathrm{eff}}$ from fitting TiO bands.\label{fig:TiO-Results}}
\end{figure}

The LTE results are hotter than the respective NLTE results at each
$\Delta T_{\mathrm{1.5D}}$ for both choices of FF, as seen in Fig. \ref{fig:TiO-Results}. 
This is not caused directly by an NLTE effect, such as Fe I overionization as in the 
previous four methods, as PHOENIX does not compute molecules in NLTE. Instead it is an 
indirect effect of the difference of LTE and NLTE $T_{\mathrm{Kin}}\left(\tau\right)$
structures in the upper atmosphere, and only 1.5D modeling of this style has 
revealed its role. For a given $T_{\mathrm{eff}}$, NLTE models are generally
hotter than LTE models in the upper atmosphere above $\tau_{12000}$ = 1, as 
seen in Fig. \ref{fig:exampleTtau}. Because of this, more TiO molecules will
collisionally dissociate in the upper atmosphere of a NLTE star than an LTE
star, and NLTE stars will form weaker absorption features. Therefore, NLTE best
fits will be required to have lower $T_{\mathrm{eff}}$ values to match
the strength of the TiO absorption features. In this case $\Delta T_{\mathrm{NLTE}}$
is non constant as a function of $\Delta T_{\mathrm{1.5D}}$, increasing
from 50 K at $\Delta T_{\mathrm{1.5D}}$ = 0 K, to a local maximum
of 100 K, then decreasing back to 75 K at maximum $\Delta T_{\mathrm{1.5D}}$,
for both FF.

\section{SUMMARY AND CONCLUSIONS \label{summary}}

The goal of this work has been to analyze the effects of massively NLTE 
atmospheric modeling combined with 2D horizontal inhomogeneities
on $T_{\mathrm{eff}}$ values inferred from SEDs and line
profiles. The stellar atmosphere and spectrum synthesis code PHOENIX
was used to generate a grid of spherical stellar atmosphere models
in both LTE and NLTE, and to synthesize spectra for the models. 

\paragraph{}

Spectra of target 2D ``observed'' stars were produced in the 1.5D
approximation by linearly averaging two NLTE 1D spectra together under
two different weighting schemes (FF), such that the $T_{\mathrm{eff,R-J}}$
was 4250 K and the temperature difference between the 1D was as large
as $\Delta T_{\mathrm{1.5D}}$ = 1050 K. The grid of LTE and NLTE 1D
SEDs and spectra were fit to the observations to infer $T_{\mathrm{eff}}$
values for the 1.5D stars using five different approaches. All inferred 
$T_{\mathrm{eff}}$ values and differences between inferred $T_{\mathrm{eff}}$ 
and 1.5D $T_{\mathrm{eff,S-B}}$ are considered to have a formal uncertainty of 
25 K, half of one temperature resolution unit in our pre-interpolated 1D grid.

\paragraph{}

Photometric colors of 1D stars computed from synthetic UBVRI photometry
were compared to 1.5D colors to assess the errors in photometrically
derived $T_{\mathrm{eff}}$ values. For the five color indices and
both values of FF, the inferred value of $T_{\mathrm{eff}}$ was seen
to increase with $\Delta T_{\mathrm{1.5D}}$, and increased at a greater
rate for indices that involved bluer wavebands. When the LTE and NLTE
results were compared, the $T_{\mathrm{eff}}$ values inferred from
fitting LTE colors were systematically higher than their NLTE counterparts.
The magnitude of $\Delta T_{\mathrm{NLTE}}$ was approximately constant
as a function of $\Delta T_{\mathrm{1.5D}}$ in all cases. The value
was largest when comparing U$_{x}$-B$_{x}$ inferred $T_{\mathrm{eff}}$
values, and decreased for redder indices. 

\paragraph{}

Absolute surface flux 1D SEDs were fit to the 1.5D SEDs to assess
how changes to the predicted bolometric flux introduced by the modeling
assumptions affect the inferred value of $T_{\mathrm{eff}}$. For
both values of FF, the inferred value of $T_{\mathrm{eff}}$ was seen
to increase with $\Delta T_{\mathrm{1.5D}}$. This approach showed
the lowest overall error of any of the full $\lambda$ distribution
fitting approaches in the inferred value of $T_{\mathrm{eff}}$; only
110 K difference between the fit $T_{\mathrm{eff}}$ value and $T_{\mathrm{eff,S-B}}$ at 
maximum $\Delta T_{\mathrm{1.5D}}$. Again, the inferred
$T_{\mathrm{eff}}$ values from fitting LTE SEDs were systematically
higher than those from fitting NLTE SEDs, and the magnitude of $\Delta T_{\mathrm{NLTE}}$
was approximately constant as a function of $\Delta T_{\mathrm{1.5D}}$.

\paragraph{}

Relative 1D SEDs normalized to the average continuum flux in a 10 $\textrm{\AA}$
window in the R-J tail were fit to the 1.5D SEDs to assess how the
modeling assumptions affect overall shape of the SED and the temperature
sensitive spectral features located in the blue and near-UV bands.
For both values of FF, the inferred value of $T_{\mathrm{eff}}$ was
seen to increase with $\Delta T_{\mathrm{1.5D}}$. The inferred $T_{\mathrm{eff}}$
values from fitting LTE SEDs were systematically higher than those
from fitting NLTE SEDs, and the magnitude of $\Delta T_{\mathrm{NLTE}}$
was approximately constant as a function of $\Delta T_{\mathrm{1.5D}}$.
These three complimentary methods (photometric colors, absolute surface flux
SEDs, and relative SEDs) show results consistent with each other.

\paragraph{}

Of the three spectrophotometric methods, the photometric colors give
both the highest and the lowest errors on the estimates of the $T_{\mathrm{eff}}$.
At maximum $\Delta T_{\mathrm{1.5D}}$, the U$_{x}$-B$_{x}$ index fitting 
returned up to 340 K higher than $T_{\mathrm{eff,S-B}}$, while the R-I index 
returned as low as 60 K less than $T_{\mathrm{eff,S-B}}$. The V-I fitting resulted in error values similar to 
that of the absolute SEDs, with fitted $T_{\mathrm{eff}}$ values as high as 90 K above $T_{\mathrm{eff,S-B}}$.
The relative SED fitting returned error values similar to the V-R
index, as high as 210 and 240 K above $T_{\mathrm{eff,S-B}}$ respectively.
Together, these results reinforce that red/near-IR photometry is more reliable for 
diagnosing $T_{\mathrm{eff}}$ in stars with significant horizontal inhomogeneity.

\paragraph{}

Continuum normalized 1D spectra spanning a wavelength distribution between $\lambda$ =
3000 and 13000 $\textrm{\AA}$ were fit to the 1.5D
spectra to assess how the modeling assumptions change the predicted
strength of spectral features. For both values of FF, the inferred
value of $T_{\mathrm{eff}}$ was seen to increase with $\Delta T_{\mathrm{1.5D}}$.
It is important to note that this result is consistent with the spectrophotometric
results, even though it is arrived at through a complimentary method.
This approach showed the highest overall error of any of the full
spectrum fitting approaches in the inferred value of $T_{\mathrm{eff}}$;
310 K above $T_{\mathrm{eff,S-B}}$ at maximum $\Delta T_{\mathrm{1.5D}}$. The inferred $T_{\mathrm{eff}}$
values from fitting LTE spectra were systematically higher than those
from fitting NLTE spectra, and the magnitude of $\Delta T_{\mathrm{NLTE}}$
was approximately constant as a function of $\Delta T_{\mathrm{1.5D}}$.

\paragraph{}

Predicted line profiles for the important 1D TiO bands spanning a
wavelength range between $\lambda$ = 5500 and 8000
$\textrm{\AA}$ were fit to the 1.5D TiO bands to assess how the strength
of molecular features found primarily in the cold component of the
1.5D stars affect the inferred value of $T_{\mathrm{eff}}$. This
was the only approach to show the inferred value of $T_{\mathrm{eff}}$
decreasing with increasing $\Delta T_{\mathrm{1.5D}}$. The rapid
nonlinear growth of molecular features with decreasing temperature
became the dominant aspect in determining the $T_{\mathrm{eff}}$
value, over the nonlinear contribution to the average flux from the
higher temperature 1.5D component. The inferred $T_{\mathrm{eff}}$
values from fitting LTE spectra were still systematically higher than
those from fitting NLTE spectra, and the magnitude of $\Delta T_{\mathrm{NLTE}}$
was approximately constant as a function of $\Delta T_{\mathrm{1.5D}}$.

\paragraph{}

In this work we have shown that the approximations of both horizontal
homogeneity and LTE introduce errors in the value of $T_{\mathrm{eff}}$
inferred from fitting quantities derived from models to observed quantities.
By assuming both horizontal homogeneity and LTE, the inferred $T_{\mathrm{eff}}$
values may differ from the $T_{\mathrm{eff,S-B}}$ of a star by
340 K or more, depending on the quantity used to infer the $T_{\mathrm{eff}}$.
Of the two values of FF, 1:1 produced hotter values of inferred $T_{\mathrm{eff}}$
in general. In simulating horizontal inhomogeneities it was seen that
the bolometric flux of a 1.5D star increased with $\Delta T_{\mathrm{1.5D}}$,
and a percentage of the flux was redistributed at bluer wavelengths.
Spectral features in general appeared to have been produced by
a star hotter than the $T_{\mathrm{eff,R-J}}$, except that strong molecular
features found in cooler stars were also present in the spectra. Furthermore,
for all five approaches, $\Delta T_{\mathrm{NLTE}}$ is approximately independent of 
$\Delta T_{\mathrm{1.5D}}$, and we conclude that the magnitude of the effect of 
NLTE on $T_{\mathrm{eff}}$ derived from fitting 1D models is approximately 
independent of the thermal contrast characterizing the degree of the horizontal 
inhomogeneity of the star being modeled, and only dependent on the observable quantity being fit.

\paragraph{}

While $\Delta T_{\mathrm{1.5D}}$ was the only parameter varied in the scope of this
study, other modeling parameters such as log $g$ and [Fe/H] are expected to have 
an impact on $T_{\mathrm{eff}}$ -- $T_{\mathrm{eff,S-B}}$ and $T_{\mathrm{eff}}$ 
-- $T_{\mathrm{eff,R-J}}$ derived in various ways for horizontally inhomogeneous 
stars. Both \citet{samadi13} and \citet{tremblay13} have shown the RMS temperature 
variations to increase with decreasing log $g$, and \citet{tremblay13} has also 
shown them to increase with decreasing [Fe/H].
Likewise, the values of the fitted $T_{\mathrm{eff}}$ -- $T_{\mathrm{eff,S-B}}$ 
at a given $\Delta T_{\mathrm{1.5D}}$ are expected to differ with the $T_{\mathrm{eff,R-J}}$
of the 1.5D stars. Investigating these parameters requires extensive additions 
to our 1D grid of models, as well as an updated Fe I atomic model, and will be explored in a future study.



\acknowledgments

We would like to thank the NSERC Discovery Grant program and Saint Mary's University's
Faculty of Graduate Studies and Research for funding this work. We would also like
to acknowledge Comupte Canada member ACENet for providing us with all computational
resources and CPU time.

\clearpage





\clearpage

\clearpage


\clearpage


\clearpage


\clearpage


\clearpage


\clearpage


\clearpage


\clearpage


\clearpage


\clearpage


\clearpage


\clearpage


\clearpage


\clearpage







\clearpage

\begin{deluxetable}{lrr}
\tablewidth{0pt}
\tablecolumns{3}
\tablecaption{List of atomic species treated in NLTE energy level
 calculations. Number of energy levels and line transitions included for
 each ionization stage are given.}
\tablehead{
\colhead{Element} & \colhead{I} & \colhead{II}
}
\startdata

H & 80/3160 & \nodata\phm{000}\\
He & 19/37\phm{00} & \nodata\phm{000} \\
Li & 57/333\phm{0} & 55/124\phm{00} \\
C & 228/1387 & \nodata\phm{000} \\
N & 252/2313 & \nodata\phm{000} \\
O & 36/66\phm{00} & \nodata\phm{000} \\
Ne & 26/37\phm{00} & \nodata\phm{000} \\
Na & 53/142\phm{0} & 35/171\phm{00} \\
Mg & 273/835\phm{0} & 72/340\phm{00} \\
Al & 111/250\phm{0} & 188/1674\phm{0} \\
Si & 329/1871 & 93/436\phm{00} \\
P & 229/903\phm{0} & 89/760\phm{00} \\
S & 146/349\phm{0} & 84/444\phm{00} \\
K & 73/210\phm{0} & 22/66\phm{000} \\
Ca & 194/1029 & 87/455\phm{00} \\
Ti & 395/5279 & 204/2399\phm{0} \\
Mn & 316/3096 & 546/7767\phm{0} \\
Fe & 494/6903 & 617/13675 \\
Co & 316/4428 & 255/2725\phm{0} \\
Ni & 153/1690 & 429/7445\phm{0} \\
\hline
\enddata
\label{tab:nlte species list}
\end{deluxetable}

\clearpage


\begin{deluxetable}{ccc}
\tablewidth{0pt}
\tablecolumns{3}
\tablecaption{Wavelength grid spacing used in computing PHOENIX model
 atmospheres and spectra}
\tablehead{
\colhead{Wavelength Range $\lambda$ ($\textrm{\AA}$)} & \colhead{Spacing $\Delta\lambda$ ($\textrm{\AA}$)} & \colhead{Mid Range Spectral Resolution $R$}
}
\startdata

3000 - \phm{0}4000 & 0.010 & 350000 \\
4000 - \phm{0}5000 & 0.013 & 346000 \\
5000 - \phm{0}6000 & 0.016 & 344000 \\
7000 - \phm{0}8000 & 0.023 & 326000 \\
8000 - 11000 & 0.027 & 352000 \\
11000 - 13000\phm{0} & 0.037 & 324000 \\
\hline
\enddata
\label{tab:Wavelength-grid-spacing}
\end{deluxetable}

\clearpage


\begin{deluxetable}{ccccccc}
\tablewidth{0pt}
\tablecolumns{3}
\tablecaption{1.5D model components for 1:1 and 2:1 filling
factors and resultant 1.5D $T_{\mathrm{eff,S-B}}$ (see Section \ref{intromodel}) 
values. All 1.5D stars are considered to have $T_{\mathrm{eff,R-J}}$ = 4250 K.}

\tablehead{
\colhead{} & \multicolumn{3}{c}{1:1 Filling Factor} & \multicolumn{3}{c}{2:1 Filling Factor}\\
\colhead{} & \colhead{Warm} & \colhead{Cool} & \colhead{} & \colhead{Warm} & \colhead{Cool} & \colhead{} \\
\colhead{} & \colhead{Component} & \colhead{Component} & \colhead{1.5D} & \colhead{Component} & \colhead{Component} & \colhead{1.5D} \\
\colhead{$\Delta T_{\mathrm{1.5D}}$ (K)} & \colhead{$T_{\mathrm{eff}}$ (K)} & \colhead{$T_{\mathrm{eff}}$ (K)} & \colhead{$T_{\mathrm{eff,S-B}}$ (K)} & \colhead{$T_{\mathrm{eff}}$ (K)} & \colhead{$T_{\mathrm{eff}}$ (K)} & \colhead{$T_{\mathrm{eff,S-B}}$ (K)}
}
\startdata
1050 & \nodata & \nodata & \nodata & 4600 & 3550 & 4330 \\
1000 & 4750 & 3750 & 4336 & \nodata & \nodata & \nodata \\
\phm{0}900 & 4700 & 3800 & 4320 & 4550 & 3650 & 4309 \\
\phm{0}800 & 4650 & 3850 & 4305 & \nodata & \nodata & \nodata \\
\phm{0}750 & \nodata & \nodata & \nodata & 4500 & 3750 & 4292 \\
\phm{0}700 & 4600 & 3900 & 4293 & \nodata & \nodata & \nodata \\
\phm{0}600 & 4550 & 3950 & 4281 & 4450 & 3850 & 4277 \\
\phm{0}500 & 4500 & 4000 & 4272 & \nodata & \nodata & \nodata \\
\phm{0}450 & \nodata & \nodata & \nodata & 4400 & 3950 & 4265 \\
\phm{0}400 & 4450 & 4050 & 4264 & \nodata & \nodata & \nodata \\
\phm{0}300 & 4400 & 4100 & 4258 & 4350 & 4050 & 4257 \\
\phm{0}200 & 4350 & 4150 & 4254 & \nodata & \nodata & \nodata \\
\phm{0}150 & \nodata & \nodata & \nodata & 4300 & 4150 & 4252 \\
\phm{0}100 & 4300 & 4200 & 4251 & \nodata & \nodata & \nodata \\
\phm{000}0 & 4250 & 4250 & 4250 & 4250 & 4250 & 4250 \\
\hline
\enddata

\tablecomments{The $T_{\mathrm{eff,S-B}}$ values are quoted to the closest K as computed.}

\label{tab:1.5D-SED}
\end{deluxetable}

\clearpage


\begin{deluxetable}{cccccccccc}
\rotate
\tablewidth{0pt}
\tablecolumns{10}
\tablecaption{Best fit $T_{\mathrm{eff}}$ results for NLTE.}

\tablehead{
\colhead{$\Delta T_{\mathrm{1.5D}}$} & U$_{x}$-B$_{x}$ & B-V & V-R & V-I & R-I & Absolute SED & Relative SED & Continuum Normalized Spectra & TiO Bands
}
\startdata
\cutinhead{$\mathbf{FF 1:1}$}
1000 & 4675 & 4650 & 4575 & 4425 & 4300 & 4450 & 4550 & 4650 & 4025 \\*
\phm{0}900 & 4625 & 4575 & 4500 & 4400 & 4275 & 4425 & 4500 & 4575 & 4025 \\*
\phm{0}800 & 4550 & 4525 & 4450 & 4375 & 4275 & 4400 & 4450 & 4525 & 4050 \\*
\phm{0}700 & 4500 & 4450 & 4400 & 4350 & 4275 & 4375 & 4425 & 4475 & 4075 \\*
\phm{0}600 & 4450 & 4400 & 4375 & 4325 & 4275 & 4350 & 4375 & 4425 & 4100 \\*
\phm{0}500 & 4375 & 4350 & 4325 & 4300 & 4275 & 4325 & 4350 & 4375 & 4125 \\*
\phm{0}400 & 4350 & 4325 & 4300 & 4275 & 4250 & 4300 & 4325 & 4350 & 4175 \\*
\phm{0}300 & 4300 & 4300 & 4275 & 4275 & 4250 & 4275 & 4300 & 4300 & 4200 \\*
\phm{0}200 & 4275 & 4275 & 4275 & 4250 & 4250 & 4275 & 4275 & 4275 & 4225 \\*
\phm{0}100 & 4250 & 4250 & 4250 & 4250 & 4250 & 4250 & 4250 & 4250 & 4250 \\*
\phm{000}0 & 4250 & 4250 & 4250 & 4250 & 4250 & 4250 & 4250 & 4250 & 4250 \\
\cutinhead{$\mathbf{FF 2:1}$}
1050 & 4575 & 4575 & 4550 & 4425 & 4300 & 4425 & 4475 & 4550 & 4000 \\*
\phm{0}900 & 4525 & 4525 & 4475 & 4375 & 4250 & 4400 & 4450 & 4500 & 4025 \\*
\phm{0}750 & 4475 & 4450 & 4425 & 4325 & 4250 & 4350 & 4400 & 4450 & 4050 \\*
\phm{0}600 & 4400 & 4375 & 4350 & 4300 & 4250 & 4325 & 4350 & 4400 & 4075 \\*
\phm{0}450 & 4350 & 4325 & 4300 & 4275 & 4250 & 4300 & 4325 & 4350 & 4125 \\*
\phm{0}300 & 4300 & 4275 & 4275 & 4275 & 4250 & 4275 & 4275 & 4300 & 4200 \\*
\phm{0}150 & 4250 & 4250 & 4250 & 4250 & 4250 & 4250 & 4250 & 4275 & 4250 \\*
\phm{000}0 & 4250 & 4250 & 4250 & 4250 & 4250 & 4250 & 4250 & 4250 & 4250 \\
\hline 
\enddata
\tablecomments{All results are in K and have $\pm$ 25 K uncertainty.}

\label{tab:Master-Results-Table-1}
\end{deluxetable}

\clearpage


\begin{deluxetable}{ccccc}
\tablewidth{0pt}
\tablecolumns{5}
\tablecaption{Same as Table \ref{tab:Master-Results-Table-1} but for LTE. Only diagnostics that are insensitive to Fe I overionization in the blue/UV wavebands are displayed.}

\tablehead{
\colhead{$\Delta T_{\mathrm{1.5D}}$} & V-R & V-I & R-I & TiO Bands
}
\startdata
\cutinhead{$\mathbf{FF 1:1}$}
1000 & 4600 & 4475 & 4325 & 4100 \\*
\phm{0}900 & 4525 & 4425 & 4325 & 4100\\*
\phm{0}800 & 4475 & 4400 & 4300 & 4125\\*
\phm{0}700 & 4425 & 4375 & 4300 & 4175\\*
\phm{0}600 & 4400 & 4350 & 4300 & 4200\\*
\phm{0}500 & 4350 & 4325 & 4300 & 4225\\*
\phm{0}400 & 4325 & 4300 & 4300 & 4250\\*
\phm{0}300 & 4300 & 4300 & 4300 & 4275\\*
\phm{0}200 & 4275 & 4300 & 4300 & 4300\\*
\phm{0}100 & 4275 & 4275 & 4300 & 4300\\*
\phm{000}0 & 4275 & 4275 & 4275 & 4300\\
\cutinhead{$\mathbf{FF 2:1}$}
1050 & 4575 & 4450 & 4325 & 4075 \\*
\phm{0}900 & 4500 & 4400 & 4300 & 4100 \\*
\phm{0}750 & 4450 & 4375 & 4275 & 4125 \\*
\phm{0}600 & 4375 & 4325 & 4275 & 4175 \\*
\phm{0}450 & 4325 & 4300 & 4300 & 4225 \\*
\phm{0}300 & 4300 & 4300 & 4300 & 4275 \\*
\phm{0}150 & 4275 & 4275 & 4300 & 4300 \\*
\phm{000}0 & 4275 & 4275 & 4275 & 4300 \\
\hline 
\enddata

\tablecomments{All results are in K and have $\pm$ 25 K uncertainty.}

\label{tab:Master-Results-Table-2}
\end{deluxetable}

\clearpage


\begin{deluxetable}{cccccccccc}
\rotate
\tablewidth{0pt}
\tablecolumns{10}
\tablecaption{Best fitting $T_{\mathrm{eff}}$ -- 1.5D $T_{\mathrm{eff,S-B}}$ for NLTE.}

\tablehead{
\colhead{$\Delta T_{\mathrm{1.5D}}$} & U$_{x}$-B$_{x}$ & B-V & V-R & V-I & R-I & Absolute SED & Relative SED & Continuum Normalized Spectra & TiO Bands
}
\startdata
\cutinhead{$\mathbf{FF 1:1}$}
1000 & 339 & 314 & 239 & 89 & -36 & 114 & 214 & 314 & -311 \\*
\phm{0}900 & 305 & 255 & 180 & 80 & -45 & 105 & 180 & 255 & -295 \\*
\phm{0}800 & 245 & 220 & 145 & 70 & -30 & 95 & 145 & 220 & -255 \\*
\phm{0}700 & 207 & 157 & 107 & 57 & -18 & 82 & 132 & 182 & -218 \\*
\phm{0}600 & 169 & 119 & 94 & 44 & -6 & 69 & 94 & 144 & -181 \\*
\phm{0}500 & 103 & 78 & 53 & 28 & 3 & 53 & 78 & 103 & -147 \\*
\phm{0}400 & 86 & 61 & 36 & 11 & -14 & 36 & 61 & 86 & -89 \\*
\phm{0}300 & 42 & 42 & 17 & 17 & -8 & 17 & 42 & 42 & -58 \\*
\phm{0}200 & 21 & 21 & 21 & -4 & -4 & 21 & 21 & 21 & -29 \\*
\phm{0}100 & -1 & -1 & -1 & -1 & -1 & -1 & -1 & -1 & -1 \\*
\phm{000}0 & 0 & 0 & 0 & 0 & 0 & 0 & 0 & 0 & 0 \\
\cutinhead{$\mathbf{FF 2:1}$}
1050 & 245 & 245 & 220 & 95 & -30 & 95 & 145 & 220 & -330 \\*
\phm{0}900 & 216 & 216 & 166 & 66 & -59 & 91 & 141 & 191 & -284 \\*
\phm{0}750 & 183 & 158 & 133 & 33 & -42 & 58 & 108 & 158 & -242 \\*
\phm{0}600 & 123 & 98 & 73 & 23 & -27 & 48 & 73 & 123 & -202 \\*
\phm{0}450 & 85 & 60 & 35 & 10 & -15 & 35 & 60 & 85 & -140 \\*
\phm{0}300 & 43 & 18 & 18 & 18 & -7 & 18 & 18 & 43 & -57 \\*
\phm{0}150 & -2 & -2 & -2 & -2 & -2 & -2 & -2 & 23 & -2 \\*
\phm{000}0 & 0 & 0 & 0 & 0 & 0 & 0 & 0 & 0 & 0 \\
\hline 
\enddata
\tablecomments{All results are in K. We choose to quote the values to the nearest K as computed, but they are understood to have $\pm$ 25 K uncertainty.}

\label{tab:Master-Difference-Table-1}
\end{deluxetable}

\clearpage


\begin{deluxetable}{ccccc}
\tablewidth{0pt}
\tablecolumns{5}
\tablecaption{Same as Table \ref{tab:Master-Difference-Table-1} but for LTE. Only diagnostics that are insensitive to Fe I overionization in the blue/UV wavebands are displayed.}

\tablehead{
\colhead{$\Delta T_{\mathrm{1.5D}}$} & V-R & V-I & R-I & TiO Bands
}
\startdata
\cutinhead{$\mathbf{FF 1:1}$}
1000 & 264 & 139 & -11 & -236 \\*
\phm{0}900 & 205 & 105 & 5 & -220\\*
\phm{0}800 & 170 & 95 & -5 & -180\\*
\phm{0}700 & 132 & 82 & 7 & -118\\*
\phm{0}600 & 119 & 69 & 19 & -81\\*
\phm{0}500 & 78 & 53 & 28 & -47\\*
\phm{0}400 & 61 & 36 & 36 & -14\\*
\phm{0}300 & 42 & 42 & 42 & 17\\*
\phm{0}200 & 21 & 46 & 46 & 46\\*
\phm{0}100 & 24 & 24 & 49 & 49\\*
\phm{000}0 & 25 & 25 & 25 & 50\\
\cutinhead{$\mathbf{FF 2:1}$}
1050 & 245 & 120 & -5 & -255 \\*
\phm{0}900 & 191 & 91 & -9 & -209 \\*
\phm{0}750 & 158 & 83 & -17 & -167 \\*
\phm{0}600 & 98 & 48 & -2 & -102 \\*
\phm{0}450 & 60 & 35 & 35 & -40 \\*
\phm{0}300 & 43 & 43 & 43 & 18 \\*
\phm{0}150 & 23 & 23 & 48 & 48 \\*
\phm{000}0 & 25 & 25 & 25 & 50 \\
\hline 
\enddata

\tablecomments{All results are in K. We choose to quote the values to the nearest K as computed, but they are understood to have $\pm$ 25 K uncertainty.}

\label{tab:Master-Difference-Table-2}
\end{deluxetable}

\clearpage


\end{document}